\documentclass[prd,showpacs,nofootinbib,preprintnumbers,preprint,floatfix]{revtex4}

\renewcommand{\theequation}{\arabic{section}.\arabic{equation}}

\usepackage{amsmath,amssymb}
\usepackage{amsfonts}
\usepackage{color}
\usepackage{dsfont}
\usepackage{graphicx}
\usepackage[caption=false]{subfig}

\definecolor{remarkcolor}{rgb}{0.15,0.6,0.0}

\def\e{{\rm{e}}}
\def\be{\begin{equation}}
\def\ee{\end{equation}}
\def\ba{\begin{eqnarray}}
\def\ea{\end{eqnarray}}
\def\dfrac{\displaystyle\frac}
\def\nn{\nonumber}
\def\lb{\label}

\def\nn{\nonumber}
\def\M{{\cal M}}
\def\A{{\cal A}}
\def\Tr{{\rm Tr}}

\begin{document}
\begin{titlepage}
\title{
    \begin{flushright}
        \begin{small}    LAPTH-043/20
        \end{small}
    \end{flushright} \vspace{1.5cm}
    Nutty Kaluza-Klein dyons revisited
}

\author{Igor Bogush}
\email{igbogush@gmail.com}
\affiliation{
    Faculty of Physics, Moscow State University, 119899, Moscow, Russia{}
}

\author{G\'erard Cl\'ement}
\email{gerard.clement@lapth.cnrs.fr}
\affiliation{
    LAPTh, Universit\'e Savoie Mont Blanc, CNRS, 9 chemin de Bellevue, \\
    BP 110, F-74941 Annecy-le-Vieux cedex, France
}

\author{Dmitri Gal'tsov}
\email{galtsov@phys.msu.ru}
\affiliation{
    Faculty of Physics, Moscow State University, 119899, Moscow, Russia
}

\author{Dmitrii Torbunov}
\email{dim-16@mail.ru}
\affiliation{
School of Physics and Astronomy, University of Minnesota Twin Cities, Minneapolis, Minnesota 55455, USA}

\begin{abstract}
We extend the previous analysis of (locally) asymptotically flat solutions of Kaluza-Klein (KK) theory by assuming that the dilaton charge is an independent parameter. This corresponds to a general nondegenerate matrix of charges within the geodesic sigma model approach and comes into contact with singular solutions of the four-dimensional Einstein-scalar theory. New features of the degenerate class of solutions, which includes regular KK black holes, are also revealed. Solving the constraint equation, we find three distinct branches of the dilaton charge as a function of the other asymptotic charges, one of which contains the previously known solutions, and the other two, related by electric/magnetic duality, are new and singular. We also investigate whether a super-extreme non-rotating solution in the presence of a Newman-Unti-Tamburino (NUT) charge can become a wormhole, as is the case in Einstein-Maxwell theory. It is shown that the dilaton prevents this possibility, while non-traversable five-dimensional vacuum gravitational wormholes can exist. Finally, we analyze the geodesic structure within the chronosphere around the Misner string of Nutty KK dyons, showing that there are no closed timelike geodesics.

\end{abstract}

 \maketitle
\end{titlepage}

\setcounter{equation}{0}
\section{Introduction}

Classical solutions  to vacuum five-dimensional gravity independent of the fifth coordinate confined to a circle (Kaluza-Klein theory) were extensively studied in the past, most notably in Refs. ~\cite{leut, Dobiasch:1981vh,Chodos:1980df,Gross:1983hb,Sorkin:1983ns,Clement:1986bt,Clement:1985gm,Gibbons:1985ac,Rasheed:1995zv}. Other important work developing  mathematical tools and studying the exact solutions includes  Refs. \cite{neuge, Maison:1979kx,BeRu80,Clement:1986dn,Frolov:1987rj, matos,Poletti:1995yq,Aliev:2008wv} and references therein.  Previous interest in this subject was  concentrated on regular black hole solutions of this theory and their relevance to supergravity/string theory \cite{Gibbons:1982ih,Gibbons:1984hy,Gibbons:1987ps,Breitenlohner:1987dg,Garfinkle:1990qj,Cvetic:1994hv,Ortin:2015hya}.
For regular black holes the dilaton charge is not an independent parameter in accordance with the famous no-scalar hair theorems.
Later, new generalizations and extensions attracted attention, such as solutions depending on the fifth coordinate with asymptotics of five-dimensional vacuum \cite{Giusto:2007fx,Niarchos:2008jc,Tomizawa:2008rh,Horowitz:2011cq},  hairy black holes in Einstein-Maxwell-scalar theories with more general coupling functions \cite{Herdeiro:2015waa,Mejias:2019aio,Astefanesei:2019pfq,Grunau:2019bsd}, solutions relevant to holography \cite{Azeyanagi:2008kb,Goldstein:2009cv} and astrophysical applications \cite{Hirschmann:2017psw,Jai-akson:2017ldo,McCarthy:2018zze}. Another aspect concerns solutions containing naked singularities: if earlier these were rejected completely as nonphysical, they have recently attracted interest for modeling the metrics of ultracompact astrophysical objects outside the Kerr paradigm  \cite{Jusufi:2018gnz}, or as sources for generating regular solutions of modified gravity \cite{Galtsov:2018xuc,BenAchour:2019fdf,Domenech:2019syf}. One particular type of singularity is the Misner string in solutions endowed with a NUT charge. It was suggested \cite{Clement:2015cxa} that in the Bonnor interpretation (as generated by some singular matter source) such solutions can be rehabilitated in a sense, and within the Einstein-Maxwell theory may give rise to new type of wormholes \cite{Clement:2015aka}.

Keeping in mind the possible physical relevance of singular solutions and solutions with NUT, we undertook a revision of more general classes of solutions of the original KK theory, in which the dilaton charge is considered as an independent parameter.
Our study is based on the sigma-model representaion of the KK theory along the line of the Refs.
~\cite{Dobiasch:1981vh,Clement:1986bt,Clement:1986dn,Rasheed:1995zv}.
In four dimensions the KK theory is equivalent to the four-dimensional Einstein-Maxwell-dilaton (EMD) theory  \cite{Galtsov:1995mb} with the dilaton coupling constant $\alpha=\sqrt{3}$.
The main generating method to construct stationary solutions for $\alpha=\sqrt{3}$ is further dimensional reduction to a three-dimensional sigma model \cite{neuge, Maison:1979kx,BeRu80,Clement:1986dn}
on the coset space   $G=SL(3,R)/SO(2,1)$. It turns out that the Einstein-Maxwell and KK theories are the only ones from the family of EMD theories with general dilaton coupling which admit a coset representation \cite{Galtsov:1995mb} (solutions for arbitrary $\alpha$ were studied numerically, in particular, in \cite{Poletti:1995yq,Galtsov:2014wxl}). To solve the sigma-model equations one can apply the geodesic approach first introduced by Kramer and Neugebauer \cite{Neugebauer:1969wr}, in which the potentials are parameterized by the so-called charge matrix $A\in SL(3,R)$. Using this approach, Rasheed~\cite{Rasheed:1995zv} found a general rotating NUT-less dyon solution.
A more detailed study of thermodynamics, dual symmetries and the study of geometry near the horizon was given in \cite{Larsen:1999pp}.

Meanwhile, some questions related to KK solution space still remain underexplored which motivates the present paper. These include the following.
First, in most of the cited papers, only NUT-less dyonic solutions were explored. Second, most of the metrics constructed by the sigma-model approach
corresponded to degenerate $A$, the $\det A=0$ constraint being considered as the cosmic censorship condition \cite{Breitenlohner:1987dg}. But, as we will see here, the relation between the regularity of the horizon and the degeneracy of the charge matrix is not so direct, the condition $\det A=0$ being  only a necessary one. The charges of the static locally asymptotically flat solutions in EMD theory to which we restrict here, include the mass $M$, the NUT parameter $N$, the electric and magnetic charges $Q,\,P$ and the dilaton charge $D$.  For this degenerate class the dilaton charge is not an independent parameter, in consistency with the no-hair theorems of Einstein-scalar theory \cite{MasoodulAlam:1993ea,Yazadjiev:2010bj,Chrusciel:2012jk}. But the EMD theory also admits solutions without electric and magnetic charges and with non-zero dilaton charge, known as Fisher-Janis-Newman-Winicour (FJNW)  solutions~\cite{Fisher:1948yn,Bergmann:1957zza,Penney:1968zz,Janis:1968zz,Abdolrahimi:2009dc}. These have a singular horizon and they attracted  wide interest recently as a simple model of naked singularities \cite{Gyu}, which could be used to describe observable properties of compact objects beyond the Kerr paradigm \cite{Jus}. The FJNW solution corresponds to a particular non-degenerate charge matrix of the static sector of EMD~\cite{Rasheed:1995zv}. But the full KK generalization of the FJNW solution was not explored so far. Here we construct the generic KK locally asymptotically flat static solution with an independent dilaton charge and explore various particular cases.

At the same time, as we show here in detail, the $\det A=0$ subclass also contains singular solutions. The charge matrix degeneracy condition is a cubic equation with respect to the dilaton charge $D$ which generically has three solutions defining three branches of KK dyons. Between them, one branch contains the previously known black holes, while the two other branches (related by electric-magnetic duality) are new and generically describe naked singularities. Another new interesting feature that we reveal here consists in   periodicity of the solution family in the space of parameters. Namely, by introducing two mixing angles for the electric/magnetic charges and mass/NUT charges, we find that the cubic constraint equation exhibits periodicity in terms of a certain linear combination of these angles, which entails periodicity of the solution properties in the parameter space.

Let us recall that electric and magnetic KK configurations are related by a discrete duality inverting the sign of dilaton. The corresponding four dimensional spacetime metrics are similar. But from the five-dimensional point of view, they are essentially different: the electric solution is singular, while the magnetic monopole is regular and corresponds to the product of the Euclidean Taub-NUT metric with time. Chodos and Detweiler~\cite{Chodos:1980df} have found another 5D-regular solution, which is purely electric in the four-dimensional interpretation and represents a five-dimensional wormhole. But it was unknown whether it admits dyonic and Nutty generalizations, and if yes, what would be their four-dimensional interpretation. On the other hand, as was recently shown in~\cite{Clement:2015aka}, the four-dimensional Brill solution, which is essentially the Reissner-Nordstrom solution of the Einstein-Maxwell theory endowed with a NUT parameter, becomes a four-dimensional wormhole in the overcharged case, in other words, the RN naked singularity is converted to a wormhole once the NUT charge is added. The question arises whether the superextremal non-rotating KK ($\alpha=\sqrt3$ EMD) black hole can similarly be converted into a wormhole. We show that the answer is negative: contrary to the Einstein-Maxwell case, in the KK theory the NUT charge does not convert the superextreme naked singularities into four-dimensional wormholes. But the five-dimensional wormholes in the KK theory are shown to exist; they generalize the Chodos-Detweiler electric solution to a more general solution with four independent parameters constrained by two inequalities.

We also investigate geodesics in the new metrics, aiming to clarify the potential troubles associated with the Misner string. Although Nutty solutions run into interpretational problems due to the presence of a chronology-violating region around the Misner string, it was shown in Ref.~\cite{Clement:2015aka} within the Einstein-Maxwell theory that there are no closed timelike geodesics. We address here the same question  within the KK theory.

 The plan of the paper is as follows.
In Section II we briefly describe the derivation of the three-dimensional $\sigma$-model and recall the associated matrix representation. In Section III we derive the asymptotically locally flat solution with free scalar charge, corresponding to the KK generalization of the Fisher solution. In Section IV we construct the solutions corresponding to a degenerate charge matrix, splitting them into three dilaton classes in Section V. Extremal solutions are classified in Section VI. Then we discuss the five-dimensional interpretation with an emphasis on wormholes (Section VII). Finally, in Section VIII we discuss the geodesic structure, in particular, inside the chronosphere around the Misner string. In Appendix A we analytically prove the absence of four-dimensional wormholes, while Appendix B is devoted to the relationship between five and four-dimensional geodesics.

\setcounter{equation}{0}
\section{Generating technique}
\label{sec:technique}
We start with   five-dimensional vacuum Einstein gravity
 \be\label{ac5}
S = \int d^5x\sqrt{|g_5|}R_5,
 \ee
assuming the existence  of a spacelike Killing vector $\partial/\partial x^5$. The standard Kaluza-Klein  ansatz reads:
 \be\label{met5}
ds_5^2 = \e^{-2\phi/\sqrt3}ds_4^2  - \e^{4\phi/\sqrt3}(dx^5+2\A_\mu dx^\mu)^2,
 \ee
where the dilaton $\phi$, the KK vector $\A_\mu$ and the four-dimensional spacetime metric depend only on $x^\mu=t, x^i$.  Integrating the action (\ref{ac5}) over the cyclic coordinate $x^5$,   adjusting gravitational constants and the compactification radius, and omitting some total derivative, one obtains the four-dimensional   Einstein-Maxwell-dilaton (EMD)  action:
\begin{equation}\label{ac4}
    \mathcal{S} = \frac{1}{16\pi}
    \int d^4x \;\sqrt{-g} \Big(
        - R + 2(\partial \phi)^2 - \e^{-2\alpha\phi} F^2
    \Big),
\end{equation}
with the dilaton coupling constant $\alpha=\sqrt3$, where $F=d\A$ is the Maxwell 2-form.

\subsection{3D $\sigma$-model}
With the assumption of a time-like Killing vector $\partial_t$, this four-dimensional theory in turn can be reduced to a three-dimensional $\sigma$ model (see details in \cite{Galtsov:1995mb}). One parameterizes the four-dimensional interval as
\begin{equation}
    ds_4^2 = f (dt - \omega)^2 - f^{-1} h_{ij} dx^i dx^j,
\end{equation}
where $f$ is a real function, $\omega = \omega_i dx^i$ is a three-dimensional 1-form, and $h_{ij}$ is the 3-metric, $i=1,2,3$, depending only on  $x^i$. Resolving part of the Maxwell equations and Bianchi identities, one introduces electric $v$ and magnetic $u$ potentials
\begin{equation}\label{eq:sigma.F}
        F^{ij} = \frac{f}{\sqrt{2}} \e^{2\alpha\phi} \epsilon^{ijk}\partial_k u,
        \qquad
        F_{i0} = \frac{1}{\sqrt{2}} \partial_i v,
\end{equation}
where $\epsilon^{ijk}=\pm (\text{det}\,h_{ij})^{-1/2}$ is the three-dimensional anti-symmetric tensor. Further, one assumes that three-dimensional indices are raised and lowered with $h^{ij}$ and $h_{ij}$. Following \cite{Israel:1972vx, Galtsov:1995mb}, one can solve the ${}^{i}_{\;\;0}$-components of the Einstein equations by introducing a twist-potential $\chi$
\begin{equation}\label{eq:sigma.chi}
    -f^2 \epsilon^{ijk} \partial_j \omega_k
    =
    v \nabla^i u - u \nabla^i v + \nabla^i \chi.
\end{equation}
The remaining equations coincide with those of a three-dimensional gravity-coupled  $\sigma$ model \cite{Galtsov:1995mb}:
\begin{equation}
    \mathcal{S}_\sigma = \int d^3x \sqrt{h} h^{ij}
   (
        \mathcal{R}_{ij} - \mathcal{G}_{AB} (\varphi) \partial_i \varphi^A \partial_j \varphi^B
   ),
\end{equation}
where $\mathcal{R}_{ij}$ is the three-dimensional Ricci tensor calculated with the metric $h_{ij}$, the target space coordinates (potentials) are $\varphi^A = \left(f, \chi, u, v, \phi \right)$ and the  target space  metric $\mathcal{G}_{AB}$ reads
\begin{equation}\label{eq:sigma.target_space}
    \mathcal{G}_{AB} d\varphi^A d\varphi^B
    = \frac{1}{2f^2} \left(df^2 + (d\chi + vdu - udv)^2\right)
    -\frac{1}{f} \left(\e^{-2\alpha\phi}dv^2 + \e^{2\alpha\phi}du^2\right)
    +2d\phi^2.
\end{equation}

The target space (\ref{eq:sigma.target_space}) with  $\alpha=\sqrt{3}$ possesses 8 Killing vectors \cite{Galtsov:1995mb}, forming the $sl(3,R)$ algebra and revealing that it is a coset space  $SL(3, R)/SO(2,1)$.

Computing the covariant derivative of the Riemann tensor of the target space metric, one finds that there are only two values of the dilaton coupling constant $\alpha=\sqrt{3},\,0$
for which the Rieman tensor is covariantly constant~\cite{Galtsov:1995mb}. The second case corresponds to the Einstein-Maxwell theory, minimally coupled to the scalar field.

\subsection{Matrix representation}
For generation purposes it is convenient to present the target space metric
in the matrix form
\begin{equation}
\mathcal{G}_{AB} d\varphi^A d\varphi^B=\frac14\Tr \left(\M^{-1}d\M \M^{-1}d\M  \right).
\end{equation}
$\M$ is a symmetric matrix belonging to the coset $SL(3, R)/SO(2,1)$. The matrix $\M$ transforms under the target space isometry as
\be
\M\to \M'=P^T\M P
\ee
with some matrix $P\in SL(3,R)$. In terms of $\M$, the sigma-model equations read
\be \lb{Meq}
\nabla_i\left(\M^{-1}\nabla^i\M \right)=0,
\ee
where $\nabla_i$ is a covariant derivative in the three-space,
and the three-dimensional Einstein equations are
\be \lb{RM}
 \mathcal{R}_{ij}=-\frac14 \Tr\left(\nabla_i \M \nabla_j \M^{-1}\right).
\ee
In terms of the above variables, the matrix representation of the coset was found in~\cite{Galtsov:1995mb}. We give it here in a slightly different form related by a similarity transformation:
\begin{equation}\lb{Ma}
    \M =
    \e^{2\alpha\phi/3}f^{-1}
    \begin{pmatrix}
        -f^2 + 2 v^2 f \e^{-2\alpha\phi} - (\chi - u v)^2
        &
        \sqrt{2} v f \e^{-2\alpha\phi} + \sqrt{2}u(\chi - u v)
        &
        \chi - u v
        \\
        \sqrt{2} v f \e^{-2\alpha\phi} + \sqrt{2}u (\chi - u v)
        &
        f \e^{-2\alpha\phi} - 2u^2
        &
        -\sqrt{2} u
        \\
        \chi - u v
        &
        -\sqrt{2} u
        &
        -1
    \end{pmatrix}.
\end{equation}

An alternative (and more familiar) derivation of the sigma-model has the advantage to directly use the $SL(2,R)$ structure of the compactification space~\cite{Maison:1979kx}.
One starts with the parametrization of the five-dimensional metric as
\begin{equation}
    ds^2_{(5)} = \lambda_{ab} (dx^a + a^a_i dx^i)(dx^b + a^b_j dx^j) + \tau^{-1} h_{ij} dx^i dx^j,
\end{equation}
\begin{equation}
    h_{ij} dx^i dx^j = dr^2 + Fr^2\left(d\theta^2 + \sin^2\theta d\varphi^2\right)
\end{equation}
with $\tau=-\det\lambda_{ab}$, $a,b = 0,5$ and $i,j=1,2,3$. Comparing with our previous ansatz we get
\begin{equation}
    \e^{-4\alpha\phi/3} = \lambda_{55},\qquad
    A_t = \lambda_{05} / 2 \lambda_{55},\qquad
    f = \tau / \sqrt{\lambda_{55}},
\end{equation}
\begin{equation}
    a^0 = -\omega d\varphi = 2N\cos\theta d\varphi,\qquad
    a^5 = 2\left( \omega A_t + A_\varphi \right) d\varphi = 2P\cos\theta d\varphi.
\end{equation}
Now we will define the dualized two-vector $V_a$:
\begin{equation}
    V_{a,i} = \tau \lambda_{ab} h_{il} \varepsilon^{ljk}a^b_{l,k}
    = \left( \lambda_{a0} N + \lambda_{a5} P \right)\sin\theta,
\end{equation}
then the matrix $\M$ will read
\begin{equation}\lb{Mb}
    \M = \frac{1}{\tau}\begin{pmatrix}
        \tau \lambda_{ab} - V_a V_b & V_a \\
        V_b & -1
    \end{pmatrix}.
\end{equation}

To solve the  equations (\ref{Meq},\ref{RM}) for non-rotating configurations, we assume, following Kramer and Neugebauer \cite{Neugebauer:1969wr}, that the target space variables $\varphi^A$ depend on the coordinates through a single scalar function $g(x^i)$, i.e., $\varphi^A(x^i)=\varphi^A\left[g(x^i)\right]$ which is a harmonic function in the three-space:
\be
\nabla^i\nabla_i g(x^i) =0.
\ee
Then $\varphi^A(g)$ will be  a geodesic in the target space, parameterized by $g$ as an affine parameter.  In the matrix form, the geodesic equation reads
\be\lb{Mge}
\frac{d}{dg}\left(\M^{-1}\frac{d\M}{dg}\right) =0.
\ee
Assuming that $g\to 0$ at spatial infinity, and denoting the value of the matrix $\M$ at $g=0$ as $\eta=\text{diag}(-1,1,-1)$, one can present the   solution of the Eq.  (\ref{Mge}) as
\be\lb{exp}
\M=\eta{\rm e}^{gA},
\ee
where $A$ is some constant matrix belonging to the Lie algebra $sl(3,R)$  satisfying the conditions
\be
A^T=\eta A \eta, \qquad \Tr A=0.
\ee
 To find $A$ explicitly, we normalize the harmonic function $g$ so that at spatial infinity
\be
g\sim \frac2{r},
\ee
and assume
an asymptotic behavior of the target space potentials
\begin{equation}
\label{eq:sigma.new_asymptote}
    f \to 1 - \frac{2M}{r},\quad
    \chi \to \frac{2N}{r},\quad
    u \to \frac{\sqrt{2}P}{r},\quad
    v \to \frac{\sqrt{2}Q}{r},\quad
    \phi \to \frac{\alpha D}{r}.
\end{equation}
We then find that the asymptotic value of $\M$ is  the constant matrix $ \eta = \text{diag}(-1,1,-1) $, while the charge matrix $A$ will be parameterized by six independent charges as follows:
 \begin{equation}
    A =  \begin{pmatrix}
        -M+D & -Q & N \\
        Q     & -2D  & P \\
        N     & -P & M+D
    \end{pmatrix}
    \ee

The isometries of the target space preserving the asymptotic conditions (\ref{eq:sigma.new_asymptote}) induce a transformation of the charge matrix $A$ of the form:
\be
A\to A'= P^{-1} A P,\qquad P^T\eta P=\eta.\ee
The new coset matrix $\M'=\eta {\rm e}^{gA'}$ will lead to a new solution with the same asymptotics. Clearly, for this to be true, the transformation matrix $P$ must belong to the isotropy subgroup $H=SO(2,1)$ of the isometry group. This is a convenient way to present transformations of the solution preserving its asymptotic form.

\setcounter{equation}{0}
\section{Solutions with independent dilaton charge}
Following the approach of \cite{Clement:1985gm}, we classify solutions according to the rank of the charge matrix $A$.
Solutions with an independent dilaton charge correspond to the rank three, i.e., non-degenerate matrix, $\det A\neq 0$.  The matrix exponential   can then be found  using the Lagrange interpolation formula
\begin{equation}
    \e^{gA} = \sum_{i=-1}^{+1} \e^{g \lambda_i} \prod_{i \neq j} \frac{A - \lambda_j}{\lambda_i - \lambda_j},
\end{equation}
where $\lambda_k$ are eigenvalues of  $A$ satisfying the equation
\begin{equation}
    \lambda_k^3 - \delta^2 \lambda_k - c = 0,
\end{equation}
where we denoted
\begin{equation}
    \delta^2 = \frac{1}{2} \text{tr}{A^2} = M^2 + N^2 + 3D^2 - P^2 - Q^2,
\end{equation}
\begin{equation}
    c = \det A = 2 D (M^2+N^2-D^2) + P^2 (D-M) + Q^2 (D+M) - 2 N P Q.
\end{equation}
In the generic case, the eigenvalues labeled by $k=\pm 1, 0$ read:
\begin{equation}
    \lambda_k = \frac{2\delta}{\sqrt{3}} \cos \phi_k ,\qquad
    \phi_k = \frac{1}{3}\text{arccos}\left(z\right) + \frac{2\pi}{3} k,\qquad
    z = \frac{3\sqrt{3}}{2} \frac{c}{\delta^{3}}.
\end{equation}
Since the matrix $A$ is traceless, one may simplify the product
\begin{equation}
    \prod_{i \neq j} \frac{A - \lambda_j}{\lambda_i - \lambda_j} =
    \frac{A - \lambda_{j_1}}{\lambda_i - \lambda_{j_1}}\frac{A - \lambda_{j_2}}{\lambda_i - \lambda_{j_2}}
    =
    \frac{A^2 - (\lambda_{j_1}+\lambda_{j_2})A + \lambda_{j_1}\lambda_{j_2}}{\lambda_i^2 - (\lambda_{j_1}+\lambda_{j_2})\lambda_i + \lambda_{j_1}\lambda_{j_2}
    }.
\end{equation}
Then using the relations between the eigenvalues
\begin{equation}
    \lambda_{-1} + \lambda_0 + \lambda_{+1} = 0, \qquad
    \lambda_{-1}^2 + \lambda_{0}^2 + \lambda_{+1}^2 = 2 \delta^2,\qquad
    \lambda_{-1} \lambda_0 \lambda_{+1} = c
\end{equation}
we obtain
\begin{equation}
    \prod_{i \neq j} \frac{A - \lambda_j}{\lambda_i - \lambda_j} =
    \frac{
        A^2 + \lambda_i A + c/\lambda_i
    }{
        2 \lambda_i^2 + c/\lambda_i
    } =
    1 + \frac{ \lambda_i }{ 2 \lambda_i^3 + c }
    \left(A^2  + \lambda_i A - 2 \lambda_i^2\right).
\end{equation}
One can notice that
\begin{equation}
    \frac{ \lambda_i }{ 2 \lambda_i^3 + c } - \frac{1}{3\lambda_i^2 - \delta^2} =
    \frac{ \lambda_i^3 - \delta^2 \lambda_i - c }{ (2 \lambda_i^3 + c)(3\lambda_i^2 - \delta^2) },
\end{equation}
where the numerator is  zero by virtue of the eigenvalue equation for $\lambda_i$. The denominator can be zero only if the spectrum of $\lambda_k$ is degenerate (and the second bracket $3\lambda_i^2-\delta^2$ is zero), or $c=0$ (this condition is necessary to make the first bracket $2\lambda_i^3+c$ equal to zero). These cases will be discussed later. So one has:
\begin{equation}
    \prod_{i \neq j} \frac{A - \lambda_j}{\lambda_i - \lambda_j} =
    1 + \frac{
         A^2  + \lambda_i A - 2 \lambda_i^2
    }{
        3 \lambda_i^2 - \delta^2
    } =
    \frac{
         A^2  + \lambda_i A + \lambda_i^2 - \delta^2
    }{
        3 \lambda_i^2 - \delta^2
    }.
\end{equation}
The matrix $\M$ can then be rewritten as
\begin{equation}
    \M = \eta \e^{Ag} = \eta g_2 + \eta A g_1 + \eta (A^2 - \delta^2)g_0,
\end{equation}
where we denoted
\begin{equation}\label{gn}
    g_n = \sum_{i=-1}^{+1} \frac{ \lambda_i^{n} \e^{g \lambda_i} }{ 3 \lambda_i^2 - \delta^2 },
\end{equation}
and the two matrix terms in the expansion read
\begin{equation}
    \eta A = \left(
        \begin{array}{ccc}
         M-D & Q & -N \\
         Q & -2 D & P \\
         -N & P & -D-M \\
        \end{array}
    \right),
\end{equation}
\begin{equation}
    \eta (A^2-\delta^2) = \left(
        \begin{array}{ccc}
         2 D^2+2 M D-P^2 & N P-(D+M) Q & P Q-2 D N \\
         N P-(D+M) Q & D^2-M^2-N^2 & -D P+M P+N Q \\
         P Q-2 D N & -D P+M P+N Q & 2 D^2-2 M D-Q^2 \\
        \end{array}
    \right).
\end{equation}
This solution provides the NUTty EMD dyonic generalization of the FJNW solution if we chose
\begin{equation}
    ds_{(3)}^2 = dr^2 + r(r-2\delta)(d\theta^2 + \sin^2\theta d\varphi^2),
\end{equation}
with the harmonic function
\begin{equation}
    g = -\frac{1}{\delta} \ln F,\qquad
    F = 1 - \frac{2\delta}{r}.
\end{equation}
The $2\times 2$ matrix  $\lambda_{ab} = \M_{ab} + V_a V_b / \tau = \M_{ab} - \M_{3a} \M_{b3} / \M_{33}$  and  $\tau=-\det \lambda_{ab}$  read explicitly:
\begin{align}\label{lam}
    \lambda_{00}& = -g_2 + (M-D)g_1 + (2D(D+M)-P^2) g_0 + \tau (-N g_1 + (-2DN + PQ) g_0)^2,\nn\\
    \lambda_{05} &= Qg_1 + (-(M+D)Q + NP)g_0
    +\nn\\&+ \tau (-N g_1 + (-2DN + PQ) g_0) (P g_1 + ((M-D)P + NQ) g_0 ),\nn\\
    \lambda_{55} &= g_2 - 2D g_1 + (D^2 - M^2 - N^2) g_0 + \tau (P g_1 + ((M-D)P + NQ) g_0 )^2,\nn\\
    \tau^{-1} &= g_2 + (M+D)g_1 - (2D(D-M)-Q^2)g_0.
\end{align}
This can be simplified using the form
\begin{align}\label{lam1}
    \lambda_{ab} = \tau
    \sum_{i,j=-1}^{+1} \frac{ \e^{g (\lambda_i+\lambda_j)} P_{ab}^{ij} }{ (3 \lambda_i^2 - \delta^2)(3 \lambda_j^2 - \delta^2) },
\end{align}
where $P_{ab}^{ij}$ are some polynomials of $\lambda_i$, $\lambda_j$ and charges. Then, the diagonal part $i=j$ turns out to be proportional to the eigenvalue equation
\begin{subequations}
\begin{equation}
    P^{ii}_{00} = (2D+\lambda_i)(c + \delta^2\lambda_i - \lambda_i^3),
\end{equation}
\begin{equation}
    P^{ii}_{05} = -Q(c + \delta^2\lambda_i - \lambda_i^3),
\end{equation}
\begin{equation}
    P^{ii}_{55} = (D-M-\lambda_i)(c + \delta^2\lambda_i - \lambda_i^3),
\end{equation}
\end{subequations}
and thus zero, while the non-diagonal part is non-zero. The non-diagonal term $i\neq j$ cannot be simplified further, but we can notice that $e^{g (\lambda_i+\lambda_j)}= \e^{-g \lambda_k}$, where $k\neq i,j$. Thus, functions $\tau^{-1}\lambda_{ab}$ are linear with respect to $\e^{-g\lambda_k}$.

When both topological charges are zero, $N=P=0$, we have $\lambda_{ab} = \M_{ab}$. The quantity $c$ can then be written as
\begin{equation}
    c = (D+M)(D+M+\delta)(D+M-\delta),
\end{equation}
and the equation on $\lambda$ can be resolved in terms of charges
\begin{equation}
    \lambda_0 = D+M, \qquad
    \lambda_{\pm1} = \frac{- D - M \pm \sqrt{(M - 3 D)^2 - 4 Q^2}}{2}.
\end{equation}

\subsection{Degenerate cubic $z = \pm 1$}
In this case the cubic equation has a degenerate spectrum $\lambda_i$, so we  have to rearrange the Lagrange formula. Expanding the eigenvalues near $z = s$, with $s=\pm 1$ in terms of a small deviation $\epsilon $, we find (changing the numeration for convenience):
\begin{equation}
    \lambda_0 = \frac{2s\delta}{\sqrt{3}}(1-2\epsilon^2/3 + \ldots),\qquad
    \lambda_{\pm} = \frac{2s\delta}{\sqrt{3}}(-1/2 \pm \epsilon +\epsilon^2/3 + \ldots).
\end{equation}
 The limiting form of the  Lagrange formula will read:
\begin{equation}
    \e^{Ag} =
    \frac{(A-\lambda )^2 }{3 \delta^2}\e^{-2 g \lambda }
    -\frac{(A+2 \lambda ) (A - 4 \lambda)}{3 \delta^2} \e^{g \lambda }
    +\frac{(A+2 \lambda ) (A - \lambda )}{3 \lambda} \e^{g \lambda } g
    ,\qquad
    \lambda = -\frac{s\delta}{\sqrt{3}}
\end{equation}

\subsection{The case $ \text{tr}{A^2}=0$}
When $\delta = 0$, one eigenvalue is real and the two others are complex conjugate:
\begin{equation}
    \lambda_k = c^{1/3} \e^{i2\pi k / 3},
\end{equation}
while the relevant combinations (\ref{gn}) remain real:
\begin{equation}
    g_n = \frac{1}{3}\sum_{i=-1}^{+1} \lambda_i^{n-2} \e^{g \lambda_i}
    =
    \frac{c^{(n-2)/3}}{3} \left(
        \e^{g c^{1/3}}
        +
        2 \e^{-g c^{1/3}/2}\cos\left(\sqrt{3}g c^{1/3}/2 +2\pi (n-2) / 3\right)
    \right),
\end{equation}
where the harmonic function is
\begin{equation}
    g = \frac{2}{r}.
\end{equation}

\subsection{ Complex eigenvalues }

If $|z| > 1$, the arc-cosine function has an imaginary value and $\lambda_0$ is purely real (with the  hyperbolic functions), while $\lambda_{\pm1}$ are mutually complex conjugate. The sum of the same expressions in the Lagrange formula gives a real value, so the solution is physical.

The super-extremal solutions with imaginary $\delta$ also correspond to  one real and two complex eigenvalues conjugate to each other. All expressions in the Lagrange formula contain the square $\delta^2$, which is real. Again, the terms with conjugate eigenvalues result in a real value. To make the function $g$ real, one can perform the shift $r \to r + \delta$:
\begin{equation}
    g \to
    \frac{-1}{\delta} \ln \left(\frac{r - \delta}{r+\delta}\right)
    =
    \frac{2}{|\delta|} \left(
          \text{arccot} \frac{r}{|\delta|}
        + \pi \theta(-r/|\delta|)
    \right).
\end{equation}
Thus, as expected, the super-extremal solutions are physical too. The  inverse cotangent functions has a discontinuity at $r=0$ which can be eliminated by choosing the correct sheet using the step function $\theta$.

\subsection{Singularities}
The four-metric would represent a regular black hole if the surface $f=0$ was a regular Killing horizon. Let us first show that the function $f$ can vanish only if $F$ vanishes, i.e. at $r=2\delta$ in the non-superextremal case. We will then show that the corresponding Killing horizon is 4-singular.

The function $f$ is of the form
\begin{equation}
    f = (\tau^{-1} \Lambda_{55} )^{-1/2}
\end{equation}
where, from the last equation (\ref{lam}) and (\ref{lam1}), $\tau^{-1}$ is a linear combination of the functions $\e^{g \lambda_i}$, and $\Lambda_{55} = \tau^{-1}\lambda_{55}$ is a linear combination of the functions $\e^{-g \lambda_i}$. Thus $f$  can be zero if and only if $\tau^{-1}$ or $\Lambda_{55}$ diverges. These functions can diverge only if $g$ tends to $\pm\infty$. For the subextremal case $g = -\delta^{-1} \ln F$, and the outermost divergence of the function $g$ occurs for $F=0$ ($r=2\delta$). For the extremal case $g=2/r$, it diverges for $r=0$. For the super-extremal case it doesn't diverge. So, $f=0$ can be satisfied only if $F=0$.

While the 5-metric is singular if $\tau^{-1} = 0$, the 4-metric is singular if the four-dimensional Ricci scalar
\begin{equation}
    R = 2 f (\partial_r \phi)^2 = \frac{3}{8} \frac{\tau \lambda_{55}^{'2}}{\lambda_{55}^{5/2}} = \frac{3}{8}f \left(\frac{\lambda'_{55}}{\lambda_{55}}\right)^2
\end{equation}
diverges.

\underline{$\delta^2 > 0$, $|z| < 1$.}
In this case all the $\lambda_k$ are real and the spectrum contains both positive and negative $\lambda_k$. The functions $\e^{g\lambda_k}$ behave for $x\to0$ ($x = r - 2\delta$) as $x^{S_k}$, where
\begin{equation} \label{eq:F}
    S_k = - \lambda_k / \delta .
\end{equation}
Assuming that the $S_k$ are ordered so that $S_{-1} < S_0 < S_{+1}$, the leading  asymptotic behaviours are $\tau^{-1} \sim x^{S_{-1}}$, $\Lambda_{55} \sim x^{-S_{1}}$, $\lambda_{55} \sim x^{S_{0}}$, resulting in
\begin{equation}
        R \sim S_{0}^2\, x^{\frac{S_{+1}-S_{-1}}{2} - 2}.
\end{equation}
As $|S_k| \leq 2/\sqrt{3}$ from (3.5), the power of $x$ is always negative. And the coefficient $S_0$ is different from zero in the non-degenerate case, so that the Ricci scalar diverges on the horizon.

The case of the saturated boundary \underline{$\left|z\right| = 1$} is similar to the one just analyzed, with the difference that in the expressions for $\tau^{-1}$ and $\lambda_{ab}$ terms logarithmic in $F$ appear.

\underline{$\delta^2 > 0$, $\left|z\right| > 1$.}
Let us denote
\begin{equation}
    y = \frac{1}{6}\ln\left(2 z \left(\sqrt{z^2-1}+z\right)-1\right),\qquad
    s_z = \text{sign}(z),
\end{equation}
and rewrite $S_k$ as
\begin{equation}
    S_k =
    - \frac{s_z}{2}\left(\cosh y \pm i\sqrt{3} \sinh y\right),\;
    s_z \cosh y.
\end{equation}

If $s_z=-1$ we can perform calculations similar to those of the previous case. Keeping only the real parts, one can show that $R \sim x^{-\frac{1}{4} \cosh y - 2}$ diverges for any $y$. For the case $s_z=+1$ we look at $\tau^{-1}$, for small $x$ this behaves as
\begin{equation}
    \tau^{-1} \sim F^{-\frac{1}{2}\cosh y } \cos\left(\frac{\sqrt{3}}{2}\sinh y \ln F + \text{const}\right).
\end{equation}
This function oscillates infinitely fast when $F$ approaches zero, introducing infinitely many zeroes (singularities of the 5-metric). Near one such zero, $\tau^{-1} \sim u$ ($u\to0$), $f \sim u^{-1/2}$ and $R \sim u^{-5/2}$, so that there are infinitely many singularities outside the event horizon.

The extremal case \underline{$\delta = 0$} with $c \neq 0$ can be considered as the limit $\left|z\right| \to \infty$. In \cite{Clement:1985gm} some 5-regular solutions belonging to this case were given; they cannot be 4-regular because of the oscillations of $\lambda_{55}$. Also, these 5-regular solutions are presumably exceptional among the class of extremal solutions, the generic case being 5-singular because of oscillations of $\tau^{-1}$.

The super-extremal solutions \underline{$\delta^2 < 0$} can be regular if they represent wormholes. Though it is natural to expect that all non-degenerate $\det A \neq 0$ cases are 4-singular, it is a hard problem to prove the absence of regular wormholes in the general case. In the following paragraph we will give some singular super-extremal examples. Examples of regular wormholes are not found. In Appendix \ref{sec:4d_wormholes} we prove the non-existence of four-dimensional wormholes among the degenerate solutions $\det A = 0$. Our conjecture is that there are no four-dimensional wormholes for all values of $\det A$.

\subsection{Examples}
We will analyze the Ricci scalar for several nondegenerate examples given in the Table \ref{tab:cases}. For the subextremal solutions one can consider the quantity $r^4 R$ as a function of $F \in [0, 1)$. The multiplier $r^4$ cannot remove singularities because the function $F$ vanishes for a positive value of $r$. Analogically, for superextremal solutions one can consider $(r^2 + |\delta|^2)^2R$ as a function of $g \in (0, \pi)$, where $0$ corresponds to $r\to+\infty$ and $\pi$ to $r\to-\infty$.

Expectedly, all of these examples are singular solutions (fig. \ref{fig:ricci}). Subextremal solutions with A.I and A.II have a singularity at $F=0$ (fig. \ref{fig:ricci_sub}). Solutions A.III (fig. \ref{fig:ricci_sub}) and E.I (fig. \ref{fig:ricci_ext}) have an infinite set of singularities in the vicinity of $F=0$ and $r=0$ correspondingly. The abscissa axis has a logarithmic scale, so only few singularities are drawn. Each tuning-fork-like curve corresponds to a compact space between two singularities. Solutions with an infinite number of singular points have $|z|>1$. Superextremal solutions are represented by B.I and B.II and contain a finite set of singular points.

\begingroup
\setlength{\tabcolsep}{10pt} 
\begin{center}
\begin{table}
    \begin{tabular}{| c | c c c c c c c c |}
    \hline
    Case & $M$ & $N$ & $D$ & $Q$ & $P$ & $c$ & $\delta^2$ & $z^2$ \\
    \hline
    \multicolumn{9}{|c|}{Subextremal}\\
    \hline
    A.I     & $1$ & $0$ & $2$             & $0$  & $1$ & $-11$ & $12$ & $121/256$ \\
    \hline
    A.II    & $1$ & $2$ & $1/2$           & $0$  & $1$ & $17/4$ & $19/4$ & $7803/6859$ \\
    \hline
    A.III    & $1$ & $0$ & $1/2$           & $1$  & $0$ & $9/4$ & $3/4$ & $81$ \\
    \hline
    \multicolumn{9}{|c|}{Extremal}\\
    \hline
    E.I     & $1$ & $1$ & $-\sqrt{2}$   & $2$ & $2$ &$-8(1+\sqrt{3})$ & $0$ & $\infty$ \\
    \hline
    \multicolumn{9}{|c|}{Superextremal}\\
    \hline
    B.I     & $1$ & $1$ & $1/2$           & $2$ & $2$ & $-9/4$ & $-21/4$ & $-81/343$ \\
    \hline
    B.II    & $1$ & $0$ & $-1$            & $0$ & $3$ & $-18$ & $-5$ & $-2187/125$ \\
    \hline
   \end{tabular}
   \caption{Examples of nondegenerate subextremal, extremal and superextremal solutions.}
   \label{tab:cases}
\end{table}
\end{center}
\endgroup

\begin{figure}[h]
    \centering
    \subfloat[][] {
        \includegraphics[width=0.31\textwidth]{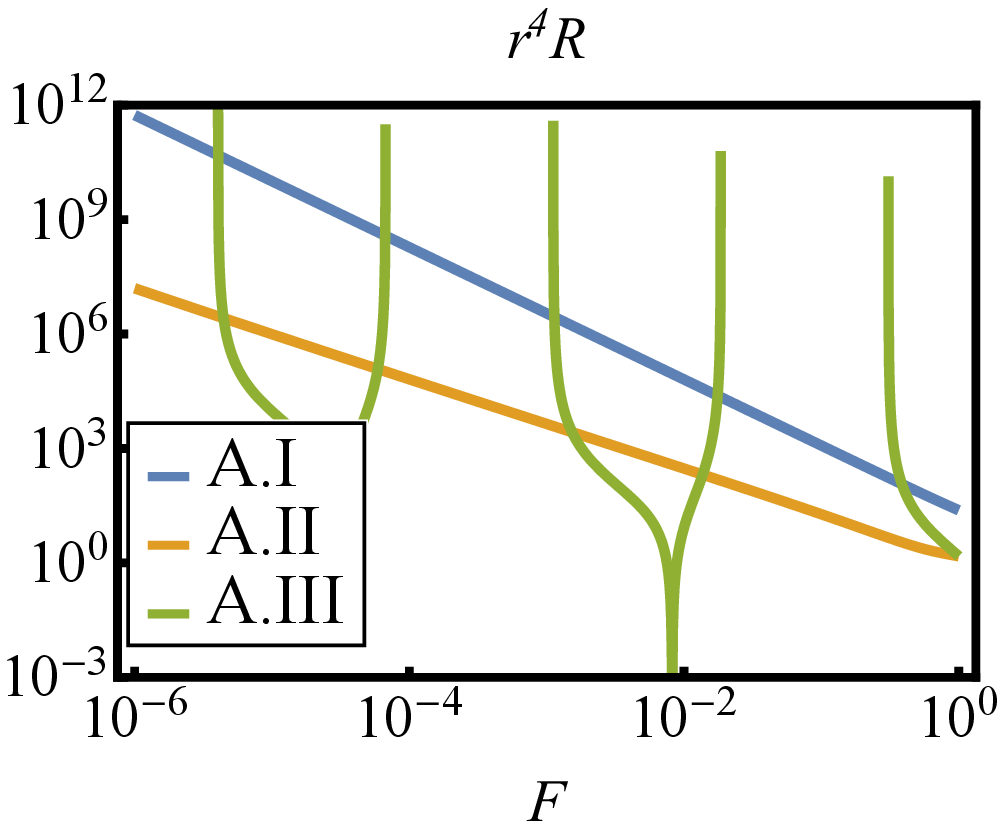}
        \label{fig:ricci_sub}
    }
    \subfloat[][] {
        \includegraphics[width=0.31\textwidth]{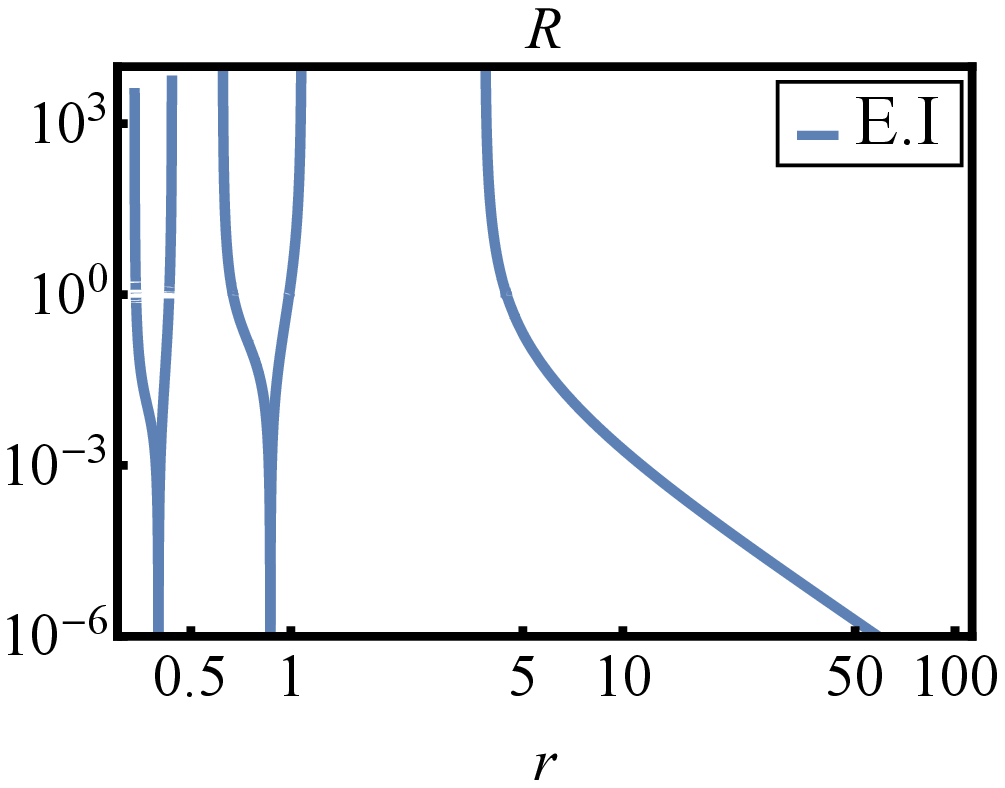}
        \label{fig:ricci_ext}
    }
    \subfloat[][] {
        \includegraphics[width=0.31\textwidth]{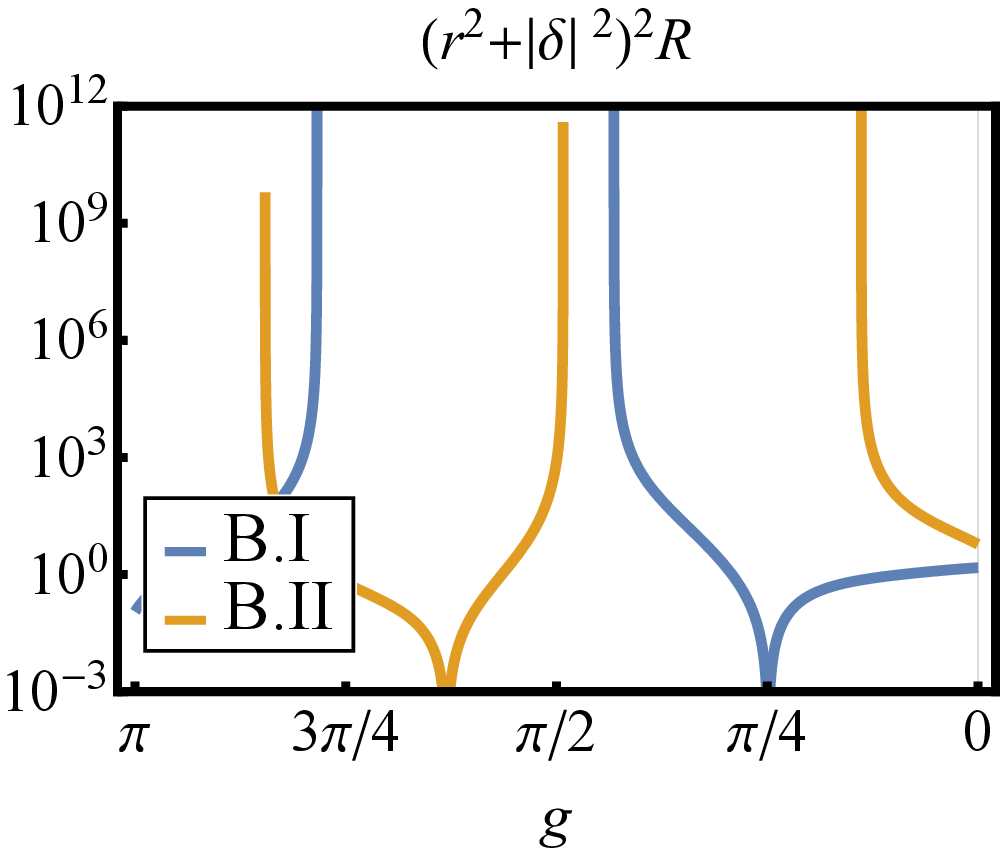}
        \label{fig:ricci_sup}
    }
    \caption{
        Ricci curvature $R$ for solutions from the Table \ref{tab:cases}.
        (\ref{fig:ricci_sub}) $r^4 R$ as a function of $F$ for subextremal solutions A.I, A.II and A.III;
        (\ref{fig:ricci_ext}) $R$ as a function of $r$ for extremal solutions E.I;
        (\ref{fig:ricci_sup}) $(r^2 + |\delta|^2)R$ as a function of $g$ for superextremal solutions B.I and B.II.
        The scale is logarithmic (except the abscissa of the fig. (\ref{fig:ricci_sup})). The spike in the lower direction corresponds to $R=0$ and is not a singularity.
        }
        \label{fig:ricci}
\end{figure}

\setcounter{equation}{0}
\section{Degenerate charge matrix}
\label{sec:solution}
It is known \cite{Clement:1986bt} that the regular black holes correspond to a degenerate charge matrix satisfying
 \be
 \det A=0,
 \ee
 which means
 \begin{equation}\label{eq:sigma.charge_constraint}
    P^2 (M - D) - Q^2 (M + D) + 2 N P Q
    = 2 D (M^2 + N^2 - D^2).
\end{equation}

In this case the general expressions (\ref{lam}) remain valid,
while the eigenvalues are
 $\lambda = -\delta, 0, +\delta$, so that
\begin{align}
    g_0 &=
    \frac{ -2 + \e^{g \delta} + \e^{-g \delta} }{ 2\delta^2 } =
    \frac{ (F - 1)^2 }{ 2\delta^2 F},\nn\\
    g_1 &=
    \frac{ \e^{g \delta} - \e^{-g \delta} }{ 2 \delta } =
    \frac{ 1 - F^2 }{ 2 \delta F},
\\
    g_2 &=
    \frac{ \e^{g \delta} + \e^{-g \delta} }{ 2 } =
    \frac{ 1 + F^2 }{ 2 F },\nn
\end{align}

Substituting $\lambda_{ab}$, $\tau^{-1}$, $g_n$, and $\delta$ into $f$, $\A_t$, and $\e^{-4\alpha\phi/3}$ and performing the shift $r \to r + \delta - M$ one finds:
\begin{equation}
    f = \frac{\Delta}{\sqrt{AB}},\qquad
    \A_t = \frac{C}{B},\qquad
    e^{-4\alpha\phi/3} = \frac{B}{A},
\end{equation}
where
\begin{subequations}
\begin{equation}
    \Delta = r(r - 2M) - 3D^2 - N^2 + P^2 + Q^2,
\end{equation}
\begin{equation}
    A = (r+D)^2 - 2D (D - M) + N^2 - P^2 + Q^2,
\end{equation}
\begin{equation}
    B =
    (r-D)^2 - 2 D (D+M) + N^2 + P^2 - Q^2,
\end{equation}
\begin{equation}
    C = Q r + D Q - N P
\end{equation}
\end{subequations}
together with the constraint (\ref{eq:sigma.charge_constraint}).
In terms of these functions the solution will read:
    \begin{align}   \label{eq:static_solution}
        &ds^2 = f ( dt - \omega d\varphi )^2 - f^{-1} \left(
            dr^2
            + \Delta \left(
                d\theta^2
                + \sin^2\theta d\varphi^2
            \right)
            \right),
   \nn\\
       & \mathcal{A} = \frac{C}{B} (dt - \omega d\varphi) + \omega_5 d\varphi,\qquad
        \e^{2\alpha\phi/3} = \sqrt{\frac{A}{B}},
   \\
        &f = \frac{\Delta}{\Sigma},
    \quad
        \Sigma = \sqrt{A B},
    \quad
        \omega = - 2 N \cos\theta,
    \quad
        \omega_5 = P \cos \theta.\nn
    \end{align}
This generalizes the solution given in \cite{Gibbons:1985ac} to include a NUT charge. The solution has outer and inner event horizons which can be found from the equation $\Delta=0$, defining two spheres with radii $r^\pm_{H}$
\begin{equation} \label{eq:solution.general.horizon}
    r^\pm_{H} = M \pm \sqrt{M^2 + N^2 + 3 D^2 - P^2 - Q^2} \equiv M \pm \delta_H.
\end{equation}
The extremal solutions (with $\delta_H=0$) will be discussed in details in section \ref{sec:extremal}.

The Eqs. $A = 0$ and $B = 0$ define up to four surfaces where the dilaton field tends to $-\infty$ and $+\infty$ respectively. These singular surfaces are spheres with radii
\begin{subequations} \label{eq:solution.general.singularity_AB}
    \begin{equation} \label{eq:solution.general.singularity_A}
    r^\pm_{A} = - D \pm \sqrt{2D(D-M)-N^2+P^2-Q^2} \equiv -D \pm \delta_A,
    \end{equation}
    \begin{equation} \label{eq:solution.general.singularity_B}
        r^\pm_{B} = D \pm \sqrt{2D(D+M)-N^2-P^2+Q^2} \equiv D \pm \delta_B.
    \end{equation}
\end{subequations}
This is confirmed by the evaluation of the scalar curvature
\begin{equation}\label{eq:solution.static.invariant_1}
    R =  -\frac{
        8 \Delta \left(
            4 N^2 - 2 \Sigma \Sigma '' +\Sigma'^2
        \right)
    }{\Sigma^3},
\end{equation}
where primes denote derivatives with respect to $r$. The Ricci scalar (\ref{eq:solution.static.invariant_1}) diverges at $\Sigma = 0$ for any set of parameters, including the case of coincident roots for both $\Delta$ and $\Sigma$. The straightforward calculation of other curvature scalars shows that $r_{A,B}^\pm$ are the only singularities.

The novel feature due to NUT is the chronology boundary given by the equation $g_{\varphi\varphi} = 0$, behind which the coordinate lines of $\varphi$ become closed timelike curves. This equation can be solved with respect to $\theta$:
\begin{equation} \label{eq:solution.general.chronology}
    \tan^2\theta = 4N^2 \frac{\Delta}{AB}.
\end{equation}
Some examples of the relative location of the above surfaces are shown\footnote{
    The figures were constructed with transformations $r = \sqrt{x^2 + y^2},\quad \theta = \arccos\left(y/r\right)$ after a shift $r \rightarrow r - \Delta r$, where $\Delta r$ is indicated for each figure. The shift $\Delta r$ was executed to reflect the whole structure of the surfaces. Black dots indicate the center of figures. Black circles are horizons. If a black circle is dashed, then it coincides with another surface. Red and blue curves are surfaces of $\phi = + \infty$ and $\phi = - \infty$ respectively. Purple curves stand for points where both $A$ and $B$ are zero, so the dilaton field is bounded. Orange curves are chronology boundaries.
} in the fig. \ref{fig:surfaces.all}. Depending on the asymptotics of the expression  $\Delta/AB$, the chronology boundary can touch the corresponding surface either at the polar axis (fig. \ref{fig:surfaces.surface_3}) or at the equator (fig. \ref{fig:surfaces.surface_1}), or intersect (fig. \ref{fig:surfaces.surface_2}).

\begin{figure}[!ht]
    \subfloat[][] {
        \includegraphics[height=0.2\textheight]{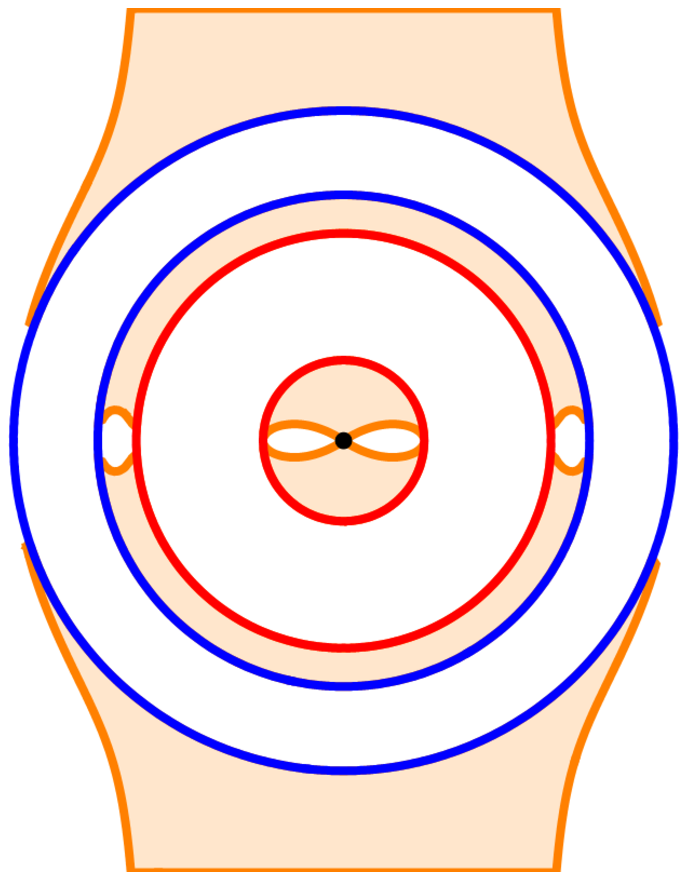}
        \label{fig:surfaces.surface_1}
    }
    \subfloat[][] {
        \includegraphics[height=0.2\textheight]{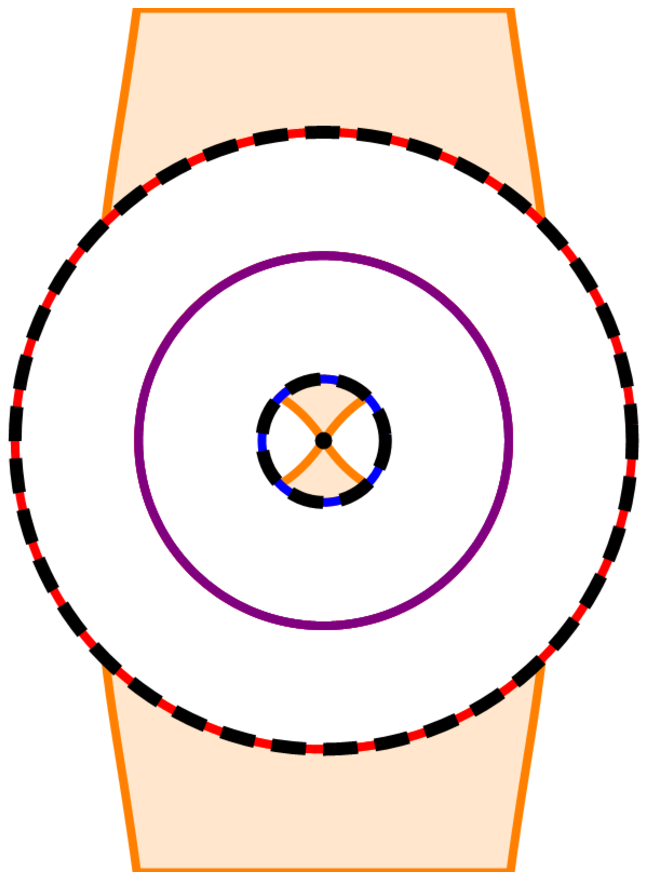}
        \label{fig:surfaces.surface_2}
    }
    \subfloat[][] {
        \includegraphics[height=0.2\textheight]{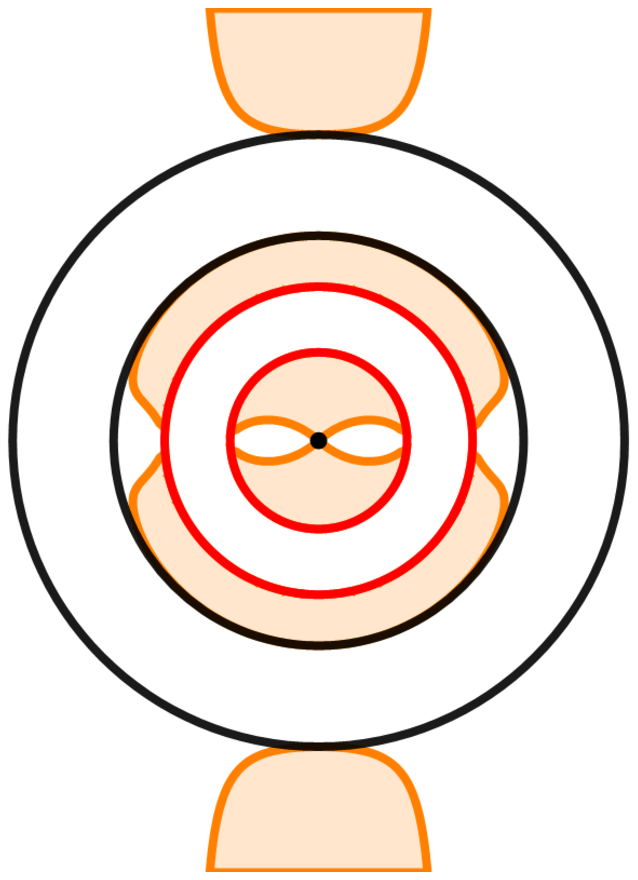}
        \label{fig:surfaces.surface_3}
    }
    \caption[]{
        Important surfaces of static solutions. Solution parameters are shown in the table \ref{tab:params1}.
    }
    \label{fig:surfaces.all}
\end{figure}
\begin{table}[!ht]
\centering
\caption{ Parameters for the solutions depicted in fig. \ref{fig:surfaces.all}. \label{tab:params1}}
\begin{tabular}{ |c||c|c|c|c|c|c| }
 \hline
 $\quad$Figure$\quad$ & $\quad M\quad $ & $\quad N\quad $ & $\quad Q\quad $ & $\quad P\quad $ & $\quad D\quad $ & $\quad \Delta r\quad $ \\
 \hline
 \hline
 \ref{fig:surfaces.surface_1} & 0.665      & -0.6 & 1.1  & 0.7 & -0.5  & 1.5 \\
 \ref{fig:surfaces.surface_2} & 0          & 1    & 0    & 0   & 1     & 3   \\
 \ref{fig:surfaces.surface_3} & $\sqrt{3}$ & 1    & 1.96 & -1  & -0.76 & 3   \\
 \hline
\end{tabular}
\end{table}
Before going to the general classification of the degenerate class of solutions, we briefly mention two well known particular cases.

\paragraph{FNJW}

~

The solution (\ref{eq:static_solution}) without NUT and with trivial Maxwell field ($N=P=Q=0$), like its counterpart in the family $\det A\neq 0$ considered in the previous section, belongs to the particular case  of FNJW  \cite{Fisher:1948yn,Abdolrahimi:2009dc}
\begin{align} \label{eq:solution.em.fisher}
    &
    ds^2 = F^S dt^2 - F^{-S}dr^2 - r^2 F^{1-S}d\Omega^2_{(2)},\qquad
    \phi = \frac{\sqrt{3}DS}{2M} \ln|F|,
    \\\nonumber &
    F = 1 - \frac{2M}{Sr},\qquad
    S = \cfrac{M}{\left(
        M^2 + 3D^2
    \right)^{1/2}},
\end{align}
with  $|S|=1/2,\,1$, corresponding to the dilaton charge $D = \pm M$ or $0$ respectively. Note that the metric functions in the case $|S|=1/2$ have only square root singularities.

\paragraph{Singly charged solutions}

~

Singly charged solutions without NUT always have horizons \cite{Gibbons:1985ac,Horne:1992bi} {because $\delta_H = M \pm D$ is always real}. Nevertheless, such solutions can be either black holes, or naked singularities, depending on whether the outermost root is that of $\Delta$ or of one of the metric functions  $A$, $B$. Here we consider the purely electric case with NUT charge ($Q \neq 0,\, P = 0$). From the constraint (\ref{eq:sigma.charge_constraint}) the electric charge is
\begin{equation}\label{eq:solution.single.charge}
    Q^2 = -2 \frac{ M^2 + N^2 - D^2 }{ M/D + 1 }.
\end{equation}
Considering (\ref{eq:solution.single.charge}), the outer roots of the functions $\Delta, A, B$ for purely electric solutions have the form (\ref{eq:solution.general.horizon}), (\ref{eq:solution.general.singularity_AB}) with
\begin{subequations}
    \begin{equation}
    \label{eq:solution.single.delta_h}
        \delta_H^2 = (M+D)^2 + N^2 \frac{M + 3D}{M + D},
    \end{equation}
    \begin{equation}
    \label{eq:solution.single.delta_a}
        \delta_A^2 = N^2\frac{D-M}{D+M},
    \end{equation}
    \begin{equation}
    \label{eq:solution.single.delta_b}
        \delta_B^2 = 4D^2 - N^2 \frac{3D+M}{D+M}.
    \end{equation}
\end{subequations}
The solution is physical and the function $\Delta(r)$ has no roots if the following two conditions hold: (\textit{i}) $Q^2 > 0$ and (\textit{ii}) $\delta_H^2 < 0$. From (\ref{eq:solution.single.delta_h}) the condition (\textit{ii}) is satisfied for $N^2 > -(M + D)^3/(M + 3D) > 0$, requiring a non-zero NUT charge. The fraction is negative if
\begin{equation}\label{eq:solution.single.int}
    1 < - M/D < 3,
\end{equation}
hence the denominator in (\ref{eq:solution.single.charge}) is negative and we require $M^2 + N^2 - D^2 > 0$ to satisfy the condition (\textit{i}). Substituting the lower bound of $N^2$ in the expression (\ref{eq:solution.single.charge}) for $Q^2$, one can get its lower bound $(Q^2)_\text{min} = 8 D^3(M + 3D)$, which is always positive under the condition (\ref{eq:solution.single.int}). Thus, for the interval (\ref{eq:solution.single.int}) and large enough $N^2$, the purely electric solution has a positive-definite  $\Delta(r)$. This could give worholes, but as we will show further, there is no static wormholes among the 4D solutions (\ref{eq:static_solution}),  which become naked singularities (contrary to the case of the Brill solution \cite{Clement:2015aka}).

With the discrete symmetry $N\to-N$, $D\to-D$ and $Q\to P$, the similar conclusions are also valid for purely magnetic solutions.

\setcounter{equation}{0}
\section{Three dilaton branches }\label{sec:classification}
Now we discuss a feature which apparently has not been sufficiently studied before, due to cubic narture to the constraint equation (\ref{eq:sigma.charge_constraint}). First note that electromagnetic and gravitational duality properties suggest the following  reparametrization of the charges:
\begin{equation} \label{eq:charge_parametrization1}
    P = e\cos\alpha\,, \;
    Q = e\sin\alpha\,, \;
    N = \mu\cos\beta\,, \;
    M = \mu\sin\beta\,,
\end{equation}
where $\alpha,\, \beta \in [0, 2\pi)$ and $e,\, \mu \geq 0$. The cubic constraint (\ref{eq:sigma.charge_constraint}) with new parameters reads
\begin{equation} \label{eq:sigma.charge_constraint_new}
  D^3 - \frac{e^2+2\mu^2}{2}D + \frac{1}{2}e^2\mu\sin(2\alpha+\beta) = 0\,.
\end{equation}
The constraint equation (\ref{eq:sigma.charge_constraint_new}) can be solved with respect to the charge $D$. All roots of the cubic equation $x^3 + px + q = 0$ are real if and only if its discriminant $\Delta_x(p,q) = -4p^3 - 27q^2$ is non-negative. Actually, the discriminant of the constraint equation  (\ref{eq:sigma.charge_constraint_new}) is
\begin{equation} \label{eq:constraint_discriminant}
   \Delta_D =
   \frac{1}{4}
   \left(e^2 - 4 \mu^2\right)^2
   \left(2 e^2 + \mu^2\right)
    + \frac{27}{4} e^4 \mu^2 \cos^2( 2\alpha + \beta ),
\end{equation}
which is a non-negative polynomial. Therefore, $D$ has three real roots (two of them can coincide when $\Delta_D=0$). Introducing new parameters
\begin{equation}
    h^2 = e^2 + 2 \mu^2\,,\qquad
    f^3 = e^2 \mu \sin(2\alpha+\beta),\qquad
    \gamma = \arccos{\frac{3^{3/2}f^3}{2^{1/2}h^3}}\,,
\end{equation}
the solution can be represented as
\begin{equation}
    D_k = \sqrt{\frac{2}{3}} h \cos\frac{\gamma - (2 k - 1)\pi}{3},
\end{equation}
where $k=-1,0,+1$. Electromagnetic duality ($\alpha \to \pi/2 - \alpha,\;\beta \to \pi - \beta,\;D\to-D$) transforms $D_{-1} \leftrightarrow D_{+1}$ into each other and $D_{0}$ into itself. As the constraint equation (\ref{eq:sigma.charge_constraint_new}) does not contain the square term, thus the sum of all roots is zero $D_{-1} + D_0 + D_{+1} = 0$.

To classify the solutions as regular black holes, singular black holes and naked timelike singularities, one should find solutions to the equations   $\delta_H = 0$, and $\delta_{A,B} = 0$, and reveal  the outermost real root among $r^\pm_{H,A,B}$.

\subsection{Domains of dilaton charge branches $D_k$}
The solutions $D_k$ for the constraint equation (\ref{eq:sigma.charge_constraint_new}) entirely belong to the regions with boundaries $\mu^2 = D^2$ (fig. \ref{fig:D_boundaries}). The branch $D_0$ satisfies the condition $D_0^2 \leq \mu^2$, while the other two branches satisfy the inequality $\pm D_{\pm 1} \geq \mu$. These branches will be shown to have different behaviour in our classification.

\begin{figure}
    \includegraphics[width=0.5\textwidth]{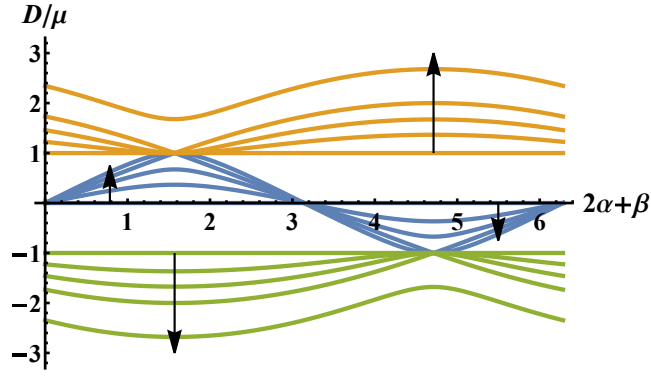}
    \caption{Values of $D_{0}/\mu$ (blue), $D_{+1}/\mu$ (brown), $D_{-1}/\mu$ (green) for different values of $e/\mu = 0, 1, 1.5, 2, 3$ as a function of $2\alpha + \beta$. Arrows demonstrate the direction of $e/\mu$ growth.}
    \label{fig:D_boundaries}
\end{figure}

The branch $D_0$ touches the other branches $D_{\pm 1}$ at the points $2\alpha + \beta = (2 \mp 1)\pi/2$ with $e=2\mu$, which allows for finding a continuous path in the charge space to connect solutions from different branches. For $e=2\mu$ the branches $D_k$ are not smooth, but they can be piece-wisely glued into smooth functions:
\begin{equation}
    \tilde{D}_k = 2 \mu \sin\left(\frac{2\alpha+\beta + 2\pi k }{3}\right),\quad k \in \mathds{Z},
\end{equation}
which follows straightforward from the constraint (\ref{eq:sigma.charge_constraint_new}).

According to these boundaries for each branch, one can suggest further developments of charge parametrization to simplify calculations. For example, the branch $D_0$ permits the following parametrization
\begin{equation}
   \mu = \sigma \cosh \gamma,\qquad
   D = \sigma \sinh \gamma,\qquad
   \sigma \geq 0,\qquad
   \gamma \in \mathds{R}.
\end{equation}
Then the constraint (\ref{eq:sigma.charge_constraint_new}) reads
\begin{equation} \label{eq:sigma.charge_constraint_new_2}
   \sigma  \left( e^2 \cosh\gamma \sin(2 \alpha + \beta) - \sinh\gamma \left( e^2 + 2 \sigma^2 \right) \right) = 0.
\end{equation}
Similar parametrizations can be introduced for the branches $D_{\pm1}$. The constraint (\ref{eq:sigma.charge_constraint_new_2}) can be easily resolved with respect to the parameter $\gamma$ in terms of inverse hyperbolic functions, which allows to write the metric functions in a simple, but lengthy way with independent parameters.

\subsection{Degenerate singularity}
If the solution is a naked singularity, the singularity can originate from the function $A$, $B$ or both of them. When the functions $A$ and $B$ have the same outer roots $r^+_A=r^+_B$, which do not coincide with their inner roots $r^-_{A,B}$, then the dilaton value at the corresponding surface is finite and only vector potential $A_\mu$ has a singularity. We will call this case a ``degenerate'' naked singularity. The equation $r_A^+ = r_B^+$ for such degenerate singularity reads
\begin{equation} \label{eq:double_singularity_1}
    2D = \delta_A - \delta_B.
\end{equation}
Squaring the equation (\ref{eq:double_singularity_1}) and using the definitions (\ref{eq:solution.general.singularity_AB}) gives
\begin{equation}
    N^2 + \delta_A \delta_B = 0,
\end{equation}
which holds if $N = 0$ and $\delta_A \delta_B = 0$. Substituting it back into (\ref{eq:double_singularity_1}) results in $\delta_A = 2D,\,\delta_B=0$ for positive $D$ and $\delta_B = 2D,\,\delta_A=0$ for negative $D$. Each of this pair of equations can be rewritten in a simple form
\begin{equation} \label{eq:double_singularity_general_1}
  -2D M + P^2 - Q^2 = 2 D|D|,
\end{equation}
Considering the constraint (\ref{eq:sigma.charge_constraint}) one can find 3 different cases with $N=0$: ${D > 0, Q = 0}$; ${D = 0, Q^2 = P^2}$; ${D < 0, P = 0}$.

In the case of a non-zero dilaton charge $D\neq 0$, the outer singularity $r^+_A=r^+_B$ necessarily coincides with one of the inner singularity $r^-_A$ or $r^-_B$, so the dilaton field is not regular at this point. These degenerate  singularities are not permitted. On the other hand, the case $D=0$ is a regular black hole, corresponding to the extremal dyonic Reissner-Nordstr\"{o}m solution.

\subsection{$D_{0}$-branch $\mu^2 \geq D^2$}
\textbf{Event horizon existence.}
The necessary condition for horizon existence is $\delta_H^2 \geq 0$. The Killing horizon becomes extremal for
\begin{equation}
        2\alpha + \beta = n \pi \pm l,
        \qquad
        l = \arcsin\left(
            \frac{
                \sqrt{e^2-\mu^2} \left(e^2+8\mu^2\right)
            }{
                3 \sqrt{3} e^2 \mu^3
            }
        \right),\qquad
        n \in \mathds{Z}.
\end{equation}
The quantity $l$ is not real for $e/\mu < 1$ and $e/\mu > 2$. One can make sure that the quantity $\delta_H^2$ is strictly positive for $e/\mu < 1$ and strictly negative for $e/\mu > 2$. The Killing horizon does not exist in the intervals $ 2\alpha + \beta \in {(n \pi - l, n \pi + l)} $ with any integer $n$ when ${1 \leq e/\mu \leq 2}$.

\textbf{Singularities.} The transformations $\alpha \to \pi/2 - \alpha$ ($P \leftrightarrow Q$) and $\beta \to \pi - \beta$ ($N \to -N$) lead to the change of the dilaton charge sign $D_{0}\to-D_{0}$ and translate the function $A$ to the function $B$ and vice verse. Therefore, results for $\delta^2_B$ can be obtained from results for $\delta^2_A$ after such  transformations. The functions $\delta^2_{A,B}$ in terms of the parameters (\ref{eq:charge_parametrization1}) are
\begin{subequations}
    \begin{equation}
        \delta^2_A =
        -\mu^2 \cos^2\beta
        + 2D_{0}(D_{0}-\mu\sin\beta)
        + e^2 \cos(2\alpha),
    \end{equation}
    \begin{equation}
        \delta^2_B =
        -\mu^2 \cos^2\beta
        + 2D_{0}(D_{0}+\mu\sin\beta)
        - e^2 \cos(2\alpha).
    \end{equation}
\end{subequations}
For the branch $D=D_{0}$, the functions $\delta^2_{A,B}$ are invariant under the transformation $\alpha \to \alpha + \pi$ or another transformation $\beta \to \beta + \pi$, which can be observed as a symmetry in the fig. \ref{fig:classification_singularity_2}. For the case $e/\mu=0$ the dilaton charge is $D_0=0$, and the singularities exist ($\delta^2_{A} \geq 0$ or $\delta^2_{B} \geq 0$) only if $\beta = \pi / 2, \; 3\pi /2$, which corresponds to the NUTless case $N=0$. Increasing the ratio $e/\mu$, these lines distend (fig. \ref{fig:classification_singularity_2} a and b). At the point $e/\mu=1$, the regions distended from the two lines touch each other and merge (fig. \ref{fig:classification_singularity_2} c and d). The regions for the values $1 < e/\mu \leq 2$, where the $A$- or $B$-singularity is absent, can be distinguished as ellipse-like, with ellipses touching each other at $\beta=\pi/2, 3\pi/2$ (fig. \ref{fig:classification_singularity_2} d, e, f). For the value $e/\mu=\sqrt{1+\sqrt{2}}$ the whole plane $(\alpha,\beta)$ is covered by regions where at least one singularity exists. This value can be found from the equation $\delta_A=\delta_B=0$ for $\beta = 0$, where the $A$- singularity touches the $B$-singularity (fig. \ref{fig:classification_singularity_2} e). For the values $e/\mu>2$ ellipse-like regions merge into vertical strips (fig. \ref{fig:classification_singularity_2} g, h). The boundaries of the strips tend to the solutions for $\mu=0$, which for $\delta^2_{A,B}=0$ is solved by $\alpha = (2n+1) \pi / 4$ with $n \in \mathds{Z}$.
    \begin{figure}[!ht]
        \includegraphics[width=1\textwidth]{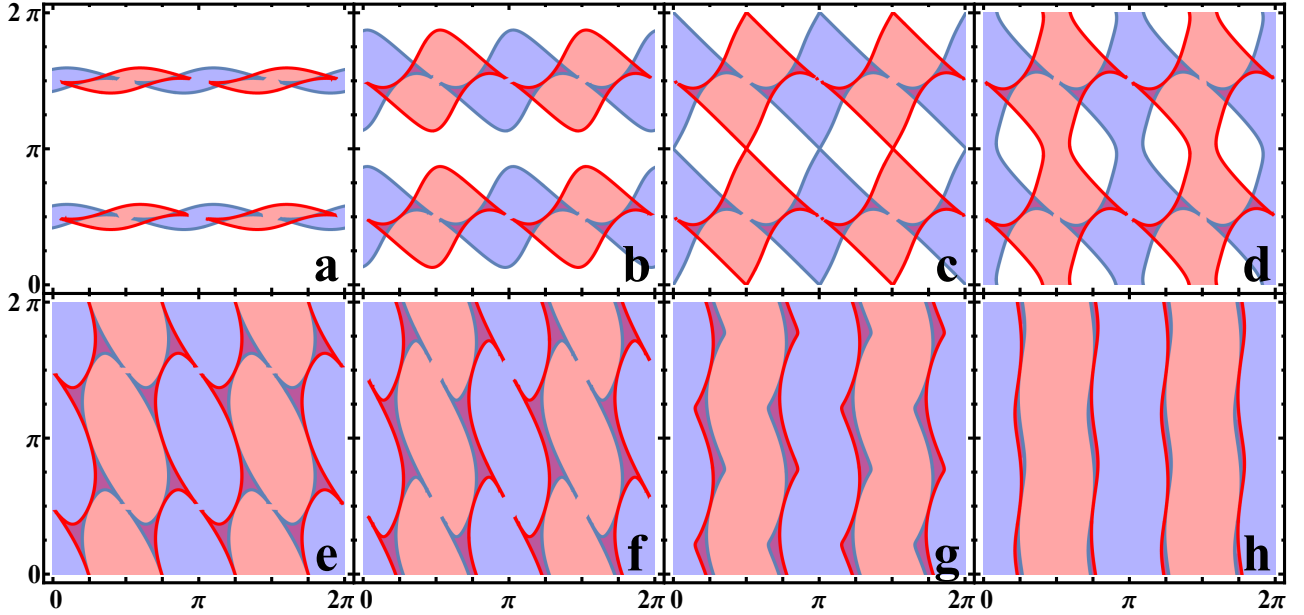}
        \caption{
            Regions of singularity existence (red for $\delta^2_B > 0$ and blue for $\delta^2_A > 0$, purple for existence of both of them) for different values $e/\mu$: (a) $0.5$, (b) $0.95$, (c) $1.0$, (d) $1.05$, (e) $\sqrt{1+\sqrt{2}}$, (f) $2.0$, (g) $2.1$, (h) $3.0$. The x- and y- coordinates stand for $\alpha$ and $\beta$ respectively. 
        }
        \label{fig:classification_singularity_2}
    \end{figure}

\textbf{Singular horizon.} A singular horizon can appear if the equation $r_A^+ = r^+_H$ or $r_B^+ = r^+_H$ holds, where the quantities at both sides should be real. Actually, such solutions may not be singular black holes due to covering of the event horizon by an another singularity. As these equations are complex enough to find their solutions analytically, they can be found numerically. The simplest case is $e=0,\,\beta=3\pi/2$ which is the Schwarzschild solution with negative mass $M$. All the other singular black holes are extremal with $r_H^+=r_A^+,\,\delta_A=0$ or $r_H^+=r_B^+,\,\delta_B=0$, i.e.
\begin{equation}
    M=-D,\qquad
    Q=\pm2M,\qquad
    P=\mp N
\end{equation}
or
\begin{equation}
    M=D,\qquad
    P=\pm2M,\qquad
    Q=\pm N.
\end{equation}
Nevertheless, not all extremal solutions are singular (fig. \ref{fig:classification_cls}). There are twice more singular black holes for $1 < e/\mu < \sqrt{2}$ than for $\sqrt{2} < e/\mu < 2$. A half of the singular black hole solutions disappear at $e/\mu = \sqrt{2}$ because a new singularity appears and covers a singular horizon.

\textbf{Classification.}
\begin{figure}
        \includegraphics[width=1\textwidth]{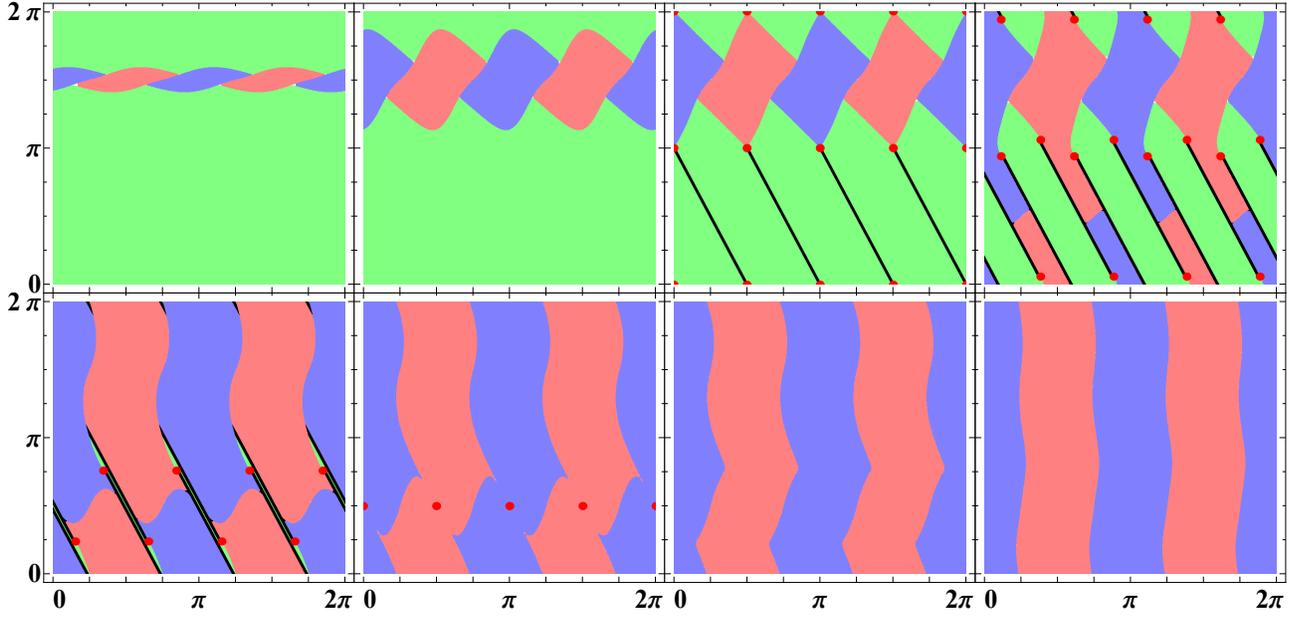}
        \caption{
            Classification of solutions (red for naked singularities coming from $r_B$, blue for naked singularities coming from $r_A$, green for black holes, black lines for extremal black holes, red dots for singular black holes) for different values $e/\mu$: (a) $0.5$, (b) $0.95$, (c) $1.0$, (d) $1.05$, (e) $\sqrt{1+\sqrt{2}}$, (f) $2.0$, (g) $2.1$, (h) $3.0$. The abscissa and ordinate stand for $\alpha$ and $\beta$ respectively.
        }
        \label{fig:classification_cls}
\end{figure}
The full classification can be found in fig. \ref{fig:classification_cls}. For $e/\mu=0$ all the plane ($\alpha,\beta$) represents a regular black hole, except the case $\beta=3\pi/2$, where it becomes a singular black hole. Increasing $e/\mu$, naked singularities appear in the vicinity of the line $\beta=3\pi/2$. There are only regular black holes for $0 \leq e/\mu \leq 1$ and positive $M$. For $1 \leq e/\mu \leq 2$, naked singularities appear in the region $M>0$ ($0<\beta<\pi$) as well. For $e/\mu\geq2$ (including $\mu = 0$) there are naked singularities only. Singular black holes appearing in $1\le e/\mu \le 2$ are extreme.

\subsection{$D_{\pm1}$-branch $\mu^2 \leq D^2$}
\begin{figure}
        \includegraphics[height=0.2\textheight]{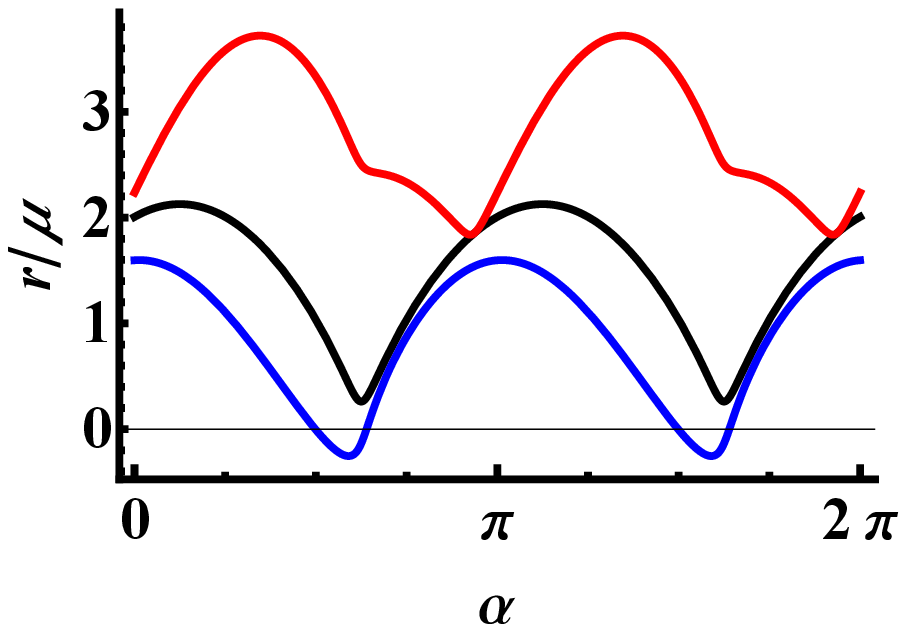}
        \includegraphics[height=0.2\textheight]{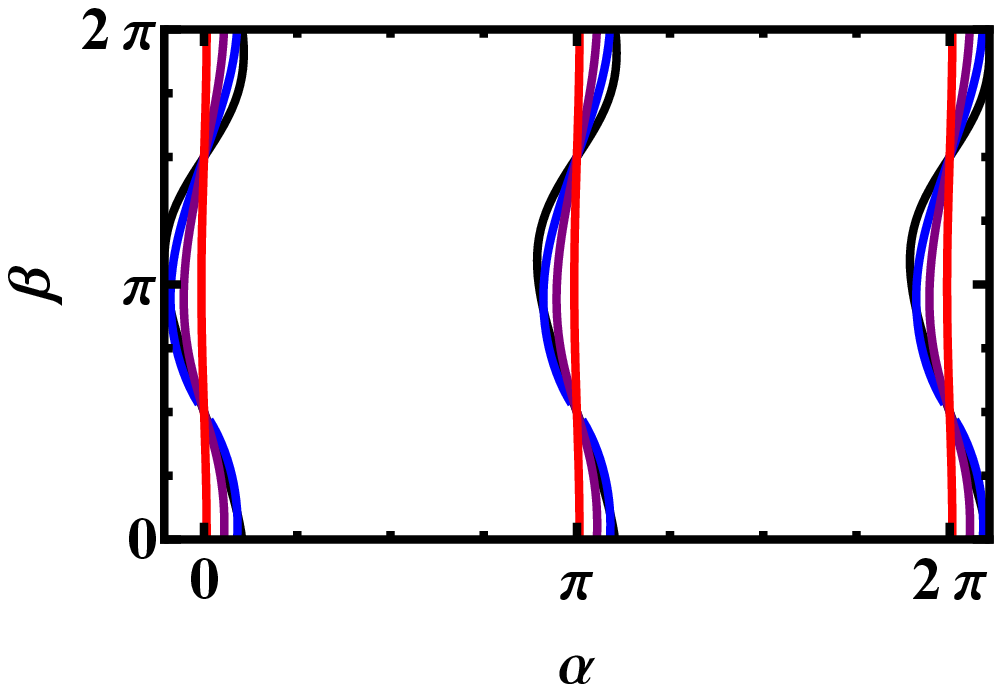}
        \caption{
            left: $r_H/\mu$ (black), $r_A/\mu$ (blue), $r_B/\mu$ (red) for $e=1.75, \beta=5\pi/4$ as a function of $\alpha$;
            right: $r_B=r_H$ for $e=0.5$ (black), $2$ (blue), $4$ (purple), $32$ (red).
        }
        \label{fig:classification_cls_2}
\end{figure}

In this section we will consider only the positive branch $D_{+1}$. To obtain results for $D_{-1}$ one should perform EM-duality and swap functions $A$ and $B$ in conclusions. As can be found numerically, the functions $\delta^2_{H,A,B}$ are always non-negative for these branches and the inequality $r_A^+ \leq r_H^+ \leq r_B^+$ always holds (fig.\ref{fig:classification_cls_2} left). Generally, these solutions are naked singularities, except particular cases of singular black holes in the vicinity of $\alpha=n\pi, n\in\mathds{Z}$ (fig.\ref{fig:classification_cls_2} right). This conclusion is consistent with the uniqueness theorem. In \cite{Yazadjiev:2010bj} Yazadjiev proved the uniqueness of asymptotically flat regular black holes (without NUT) with respect to charges $M, Q, P$ and rotation $J$. We can expect that static black hole solutions with NUT charge should be uniquely defined by the four charges $M, N, P, Q$. Despite the existence of three distinct roots $D_k$ of the charge constraint (\ref{eq:sigma.charge_constraint}), only one of them $D_0$ can be the regular black hole solution, while the two others $D_{\pm1}$ represent either a singular black hole or a naked singularity.

\setcounter{equation}{0}
\section{Extremal solutions} \label{sec:extremal}

The condition of extremal horizon $\delta_H=0$ reads
\begin{equation} \label{eq:cases.extreme.horizon_1}
3 D^2 = P^2 + Q^2 - M^2 - N^2.
\end{equation}
Substituting $D$ from (\ref{eq:cases.extreme.horizon_1}) into the charge constraint (\ref{eq:sigma.charge_constraint}) we get
\begin{eqnarray}\label{eq:cases.extreme.horizon_2}
& 27 \left[M \left(P^2-Q^2\right)+2 N P Q\right]^2 = \left(-M^2-N^2+P^2+Q^2\right) \left(8 M^2+8 N^2+P^2+Q^2 \right)^2
\end{eqnarray}
The equation (\ref{eq:cases.extreme.horizon_2}) is more compact in terms of the parameters (\ref{eq:charge_parametrization1})
\begin{equation}\label{eq:cases.extreme.horizon_4}
    (e^2 - 4\mu^2)^3 + 27 e^4\mu^2 \cos^2\left(2\alpha + \beta\right) = 0.
\end{equation}

Additionally, there is the condition of a real dilaton field $D^2 \geq 0$. From the condition of extreme horizon (\ref{eq:cases.extreme.horizon_1}), this corresponds to $3D^2 = P^2 + Q^2 - M^2 - N^2 \geq 0$, i.e. $e \geq \mu$. At the same time, the equation (\ref{eq:cases.extreme.horizon_4}) has no real solutions for $e < \mu$ and $e > 2\mu$. Thus, any solution of the equation (\ref{eq:cases.extreme.horizon_4}) leads to a real dilaton charge $D$.

It can be checked that the solution of this equation is
\begin{eqnarray}\label{eq:cases.extreme.static.condition_2}
    \left(\frac{2\mu}{e}\right)^{2/3} =
    \sin^{2/3} \left(
            \alpha +\frac{\beta}{2} -\frac{3\pi}{4}
        \right)
    +
    \cos^{2/3} \left(
            \alpha +\frac{\beta}{2}-\frac{3\pi}{4}
        \right)
\end{eqnarray}
Multiplying by $(2\mu e^2)^{1/3}$ and rewriting the trigonometric functions in terms of the initial charges, one can find
\begin{eqnarray} \label{eq:cases.extreme.static.condition_3}
    &2\sqrt{M^2 + N^2}
    = \\\nonumber & =
    \left[
        (P^2+Q^2)\sqrt{M^2 + N^2}
        + (P^2 - Q^2) M
        + 2 P Q N
    \right]^{1/3}
    +\\\nonumber&+
    \left[
        (P^2+Q^2)\sqrt{M^2 + N^2}
        + (Q^2 - P^2) M
         - 2 P Q N
    \right]^{1/3}
\end{eqnarray}
This solution is represented in the fig. \ref{fig:surfaces.extreme_static}. Note that these extreme solutions are not necessary extreme black holes, as the extreme horizon may be covered by a naked singularity or coincide with a singularity.

For $N = 0$ (fig. \ref{fig:surfaces.extreme_NUTless_static}) the equation (\ref{eq:cases.extreme.static.condition_3}) takes the form $(2M)^{2/3} = P^{2/3} + Q^{2/3}$ which agrees with Rasheed's result \cite{Rasheed:1995zv}.

\begin{figure}

    \subfloat[][] {
        \includegraphics[width=0.55\textwidth]{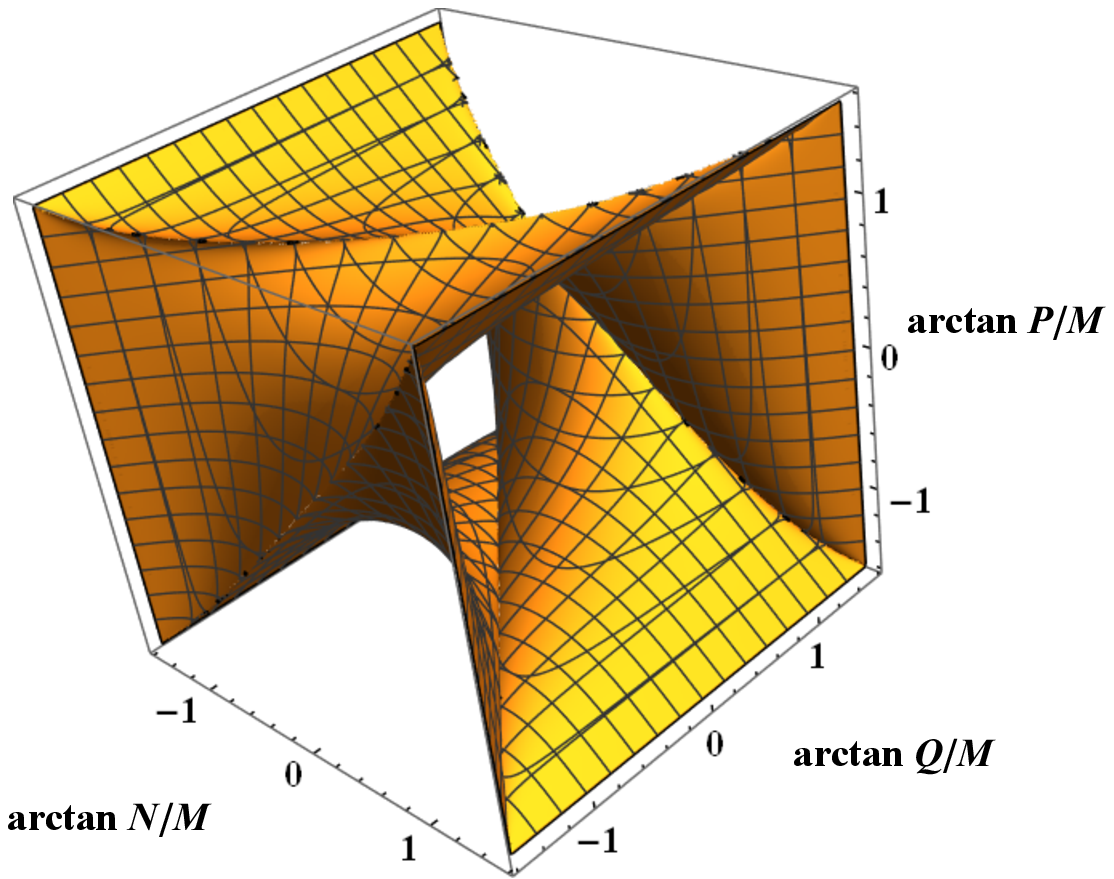}\label{fig:surfaces.extreme_static_1}
    }
    \subfloat[][] {
        \includegraphics[width=0.4\textwidth]{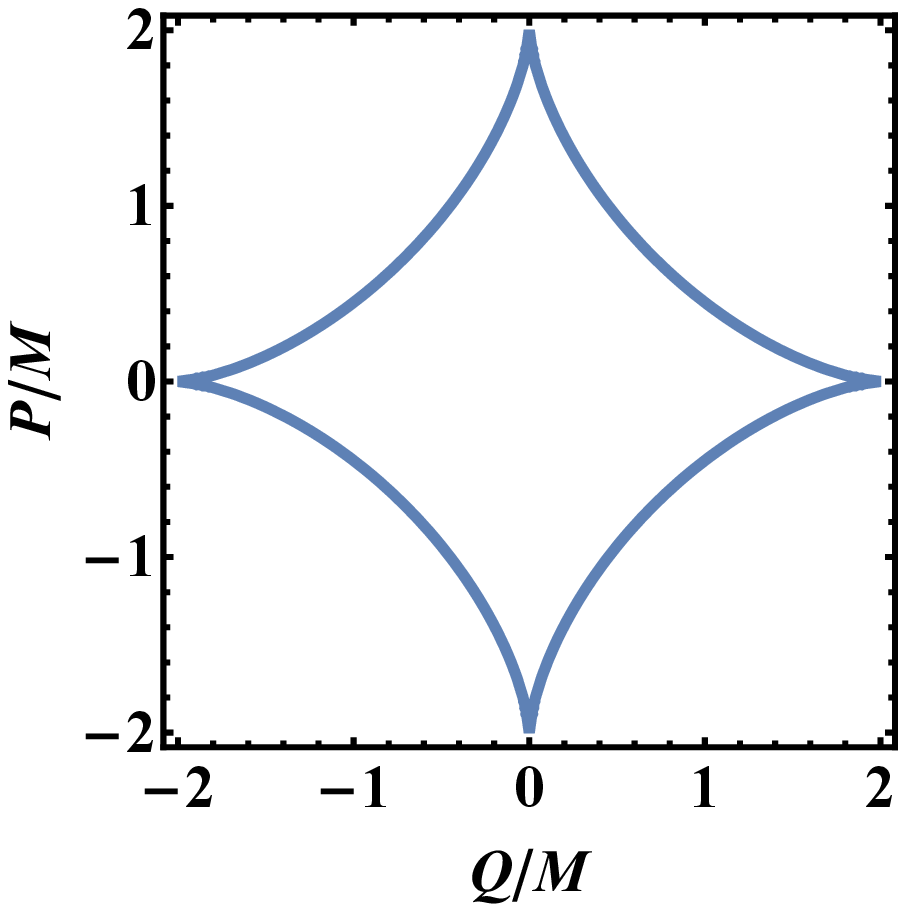}\label{fig:surfaces.extreme_NUTless_static}
    }
    \caption{Extreme static solutions in the charge space (\ref{fig:surfaces.extreme_static_1}) and extreme NUTless static solutions (\ref{fig:surfaces.extreme_NUTless_static}).}
    \label{fig:surfaces.extreme_static}
\end{figure}

\setcounter{equation}{0}
\section{Oxidation to five dimensions}\label{sec:5d}

As already mentioned, solutions of EMD theory with $\alpha^2 = 3$ can be oxidized through (\ref{met5}) to solutions of five-dimensional vacuum Einstein gravity periodic in the fifth dimension, i.e. Kaluza-Klein theory. We will only discuss here the degenerate case $\det\,A = 0$. The non-rotating stationary five-dimensional metric is such that ${\rm det}\,g = A^2\sin^2\theta$, with metric elements ($x^0 = t$)
\begin{subequations}
 \ba
&&g_{55} = -\dfrac{B}A\,, \quad g_{05} = -\dfrac{2C}A\,, \quad g_{00}
= \dfrac{F}{A} \,, \\ && g_{5\varphi} = -\dfrac{2B}A\omega_5 +
\dfrac{2C}A\omega_t \,, \quad g_{0\varphi} = -\dfrac{4C}A\omega_5 -
\dfrac{F}{A}\omega_t\,, \\ && g_{\varphi\varphi} =
-\dfrac{4B}A\omega_5^2 + \dfrac{8C}A\omega_5\omega_t +
\dfrac{F}{A}\omega_t^2 - A\sin^2\theta \,,
 \ea
\end{subequations}
where
 \be
F = \dfrac{A\Delta-4C^2}B\,,
 \ee
with an apparent singularity for $B=0$. However, after taking (\ref{eq:sigma.charge_constraint}) into account,
one obtains for the solution (\ref{eq:static_solution})
 \be \label{eq:oxidation.F}
F = (r+2D-M)^2 - M^2 - N^2 + D^2 - Q^2 - P^2\,,
 \ee
so that (as conjectured by Chen \cite{Chen:2000yi}), the 5D metric is
singularity-free provided $A>0$ for all real $r$.

\subsection{Wormholes}
If furthermore $\Delta>0$ for all real $r$, the five-dimensional metric does not correspond to a black hole,
but to a five-dimensional wormhole. So the two conditions for the existence of non-rotating five-dimensional wormholes are
 \ba
-\,\delta_A^2 &=& Q^2 - P^2 + N^2 + 2MD - 2D^2 > 0 \,, \lb{A} \\
-\,\delta_H^2 &=& Q^2 + P^2 - N^2 - M^2 - 3D^2 > 0 \,. \lb{Da}
 \ea
Adding the two together one obtains $2Q^2 > (M-D)^2 + 4D^2$, so that
$Q\neq0$ is a necessary condition for the existence of five-dimensional wormholes. The
Chodos-Detweiler wormhole \cite{Chodos:1980df} has only electric charge $Q$,
however it is not traversable \cite{AzregAinou:1990zp}. This was generalized by
Chen \cite{Chen:2000yi} to a dyonic wormhole ($M=N=D=0$) with $Q^2>P^2$ on account of
(\ref{A}).

Using the parametrization (\ref{eq:charge_parametrization1}) the constraint (\ref{eq:sigma.charge_constraint_new}) may be used to eliminate
$\sin(2\alpha+\beta)$ in terms of the three charges $e,\mu,D$,
while the two wormhole conditions (\ref{A}) and (\ref{Da}) read
 \ba
&& -e^2\cos{2\alpha} + \mu^2\cos^2\beta + 2D\mu\sin\beta - 2D^2 >
0\,, \lb{A1} \\
&& e^2 - \mu^2 - 3D^2 > 0\,. \lb{Da1}
 \ea
Then, the condition $\sin^2(2\alpha+\beta)\le1$ is equivalent to
 \be\lb{second}
(\mu^2-D^2)[(e^2-2D^2)^2-4\mu^2D^2] \ge 0\,.
 \ee
The term in square brackets is strictly positive because
 \be
2D(D\pm\mu) \le \mu^2 + 3D^2 < e^2
 \ee
on account of (\ref{Da1}). Therefore the secondary condition
(\ref{second}) is equivalent to
 \be\lb{second1}
\mu^2 \ge D^2\,.
 \ee
This condition is satisfied by the branch $D_0$.

The general analysis of the system (\ref{eq:sigma.charge_constraint_new}), (\ref{A1}), (\ref{Da1}) is complicated by the cubic character of the constraint
(\ref{eq:sigma.charge_constraint}). However the numerical investigation of the $D_0$-branch of the solution leads to the classification of five-dimensional solutions summarized in Fig. \ref{fig:classification_cls_5d}. For $e<\mu$, solutions with positive mass $M>0$ ($0<\beta<\pi$) are black holes,
while those with negative mass can be either black holes or naked singularities. For $\mu \leq e \leq 2\mu$ the solution can be a wormhole, a black hole
(either with regular or singular horizon) or a naked singularity. Solutions with $e > 2\mu$ can represent naked singularities or wormholes,
depending on both angles, but purely electric solutions are wormholes and purely magnetic ones are naked singularities.

    \begin{figure}
        \includegraphics[width=1\textwidth]{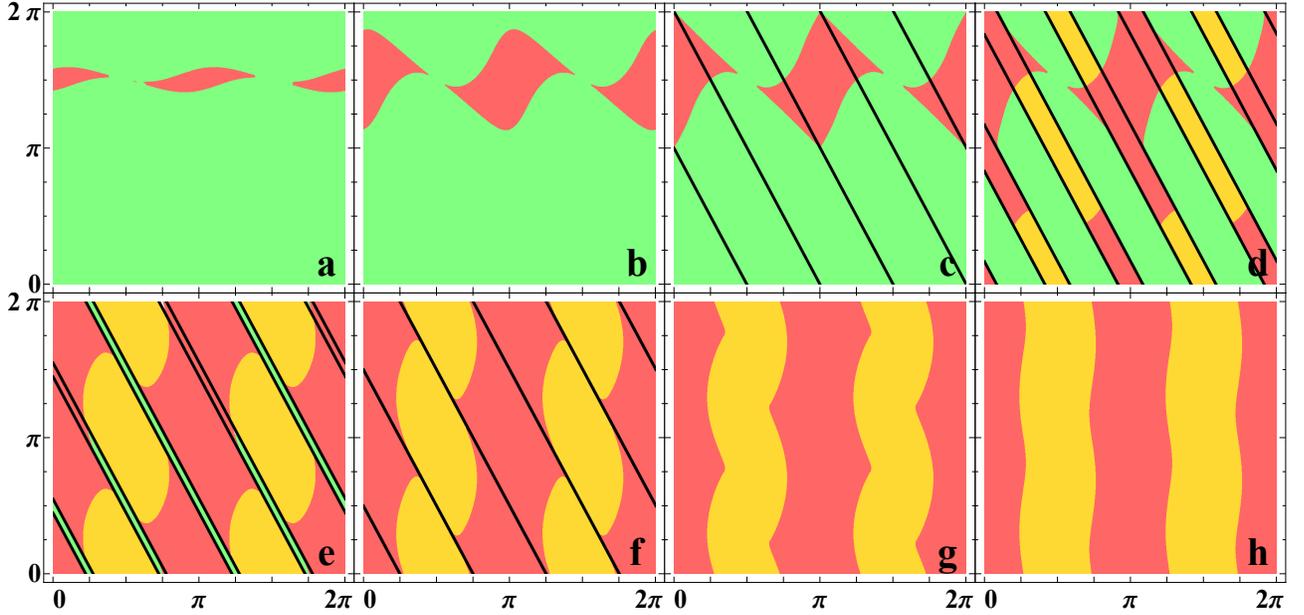}
        \caption{
            Classification of solutions in 5D for different values $e/\mu$: (a) $0.5$, (b) $0.95$, (c) $1.0$, (d) $1.05$, (e) $\sqrt{1+\sqrt{2}}$, (f) $2.0$, (g) $2.1$, (h) $3.0$. Green for black holes, red for naked singularities and yellow for wormholes. Blue lines satisfy $\delta_A^2=0$ and black lines satisfy $\delta^2_H=0$. The $x$ and $y$ axis stand for $\alpha$ and $\beta$ respectively. 
        }
        \label{fig:classification_cls_5d}
    \end{figure}

From the analysis of the 4D case, the inequality $r_B \geq r_\Delta \geq r_A$ always holds for the branch $D_{+1}$ and $r_A \geq r_\Delta \geq r_B$ for $D_{-1}$. Therefore, $D_{+1}$ and $D_{-1}$ are black holes and naked singularities in 5D respectively, consistent with (\ref{second1}).


\subsection{Special cases}

Two special cases are particularly simple:

1) \underline{$D=0$}. Then, if $\mu\neq0$ the constraint
(\ref{eq:sigma.charge_constraint_new}) is solved by $\beta=-2\alpha \, (\text{mod} \, \pi)$, with
$\cos2\alpha < 0$ from (\ref{A1}) (using $\mu^2<e^2$ from
(\ref{Da1})). In terms of the five charges $M,N,D,Q,P$, this
solution, such that
 \be
Q^2 > P^2\,, \;\; M = 2\lambda QP\,, \;\; N = \lambda(Q^2-P^2)\,,
\;\; D = 0\,,
 \ee
with $\lambda$ real bounded by
 \be
\lambda^2(Q^2+P^2) < 1\,,
 \ee
is a generalisation of Chen's dyonic wormhole.

2) \underline{$D^2=\mu^2$} ($D=\pm\mu$), which saturates the bound
(\ref{second1}). Then (\ref{eq:sigma.charge_constraint_new}) is solved by $\beta=-2\alpha
\pm \pi/2$. In terms of the original charges,
 \be
M = \lambda(Q^2-P^2)\,, \;\; N = -2\lambda QP\,, \;\; D =
-\lambda(Q^2+P^2)\,.
 \ee
The bounds (\ref{A}) and (\ref{Da}) lead to simple equations when
expressed in terms of $M$ and $D$, or
 \be
x=\lambda M = - \dfrac{MD}{Q^2+P^2}\,, \quad y = -\lambda D =
\dfrac{D^2}{Q^2+P^2}\,.
 \ee
These are respectively
 \ba
x - (x+y)^2 &>& 0\,, \\
y - 4y^2 &>& 0\,,
 \ea
and are solved by
 \be
0 < \dfrac{1-2y-\sqrt{1-4y}}2 < x < \dfrac{1-2y+\sqrt{1-4y}}2\,,
\quad 0 < y < \dfrac14\,.
 \ee
Note that $x>0$ implies $Q^2>P^2$. A simple subcase is $x=y$,
leading to
 \be
P = 0\,,\;\; N = 0\,,\;\; M = - D\,,\;\; Q^2 > 4M^2\,.
 \ee
One can directly check that this is the only solution of the system
(\ref{eq:sigma.charge_constraint_new}), (\ref{A}), (\ref{Da}) with $P=N=0$. This massive
generalization of the Chodos-Detweiler wormhole (which is recovered
for $M=0$) was previously discussed in \cite{AzregAinou:1990zp}, where it was
shown to be non traversable. The CD wormhole was shown in \cite{AzregAinou:1999}
to be unstable under small radial perturbations, the stability
status of its massive generalizations is undecided.

\subsection{Chronology boundary and ergo-region}

The chronology boundary $g_{\varphi\varphi}=0$ is the solution of
\begin{equation}
    \tan^2\theta = -4\frac{ P^2 B + 4PNC - N^2F}{A^2}.
\end{equation}
Asymptotically for $r\to\infty$ one can find
\begin{equation}\label{eq:chronology_5d}
    \tan^2\theta \approx
    \frac{4 \left(N^2-P^2\right)}{r^2}
    -\frac{8 \left(P (2 N Q - 3 D P) + M N^2\right)}{r^3}
    +\mathcal{O}(r^{-4}).
\end{equation}
If $N^2 > P^2$ the chronology boundary envelops the polar axis at the infinity with radius $\rho_s = r\sin\theta \approx 2\sqrt{N^2-P^2}$.
For the case $N^2<P^2$ the equation (\ref{eq:chronology_5d}) does not have real roots for large $r$, so the chronology boundary is compact.
One can discover an interesting case $N^2=P^2$, where the radius of the chronology boundary tends to zero at infinity. Furthermore,
we can impose the condition of extremal horizon (\ref{eq:cases.extreme.horizon_1}), and specialize to  $P =\pm N$, $Q=\mp 2M$, $D=-M$.
After transformations $r\to r + M$ and $t \to t \pm x^5$ this solution has the form
\begin{equation}
    ds^2 = \left(1-\frac{4M}r - \frac{2N^2}{r^2}\right)dt^2 \pm 2dt(dx^5 \pm 2N\cos\theta d\varphi)dt - dr^2 - r^2 (d\theta^2 + \sin^2\theta d\varphi^2).
\end{equation}
We recognize in this extremal metric the one-center case of a multi-center five-dimensional metric constructed in \cite{Clement:1985gm} (Eq. (49)).
It represents an electric monopole endowed with NUT charge. Remarkably, although two Dirac-Misner strings extend along the symmetry axis from the source $r=0$
to $\pm\infty$, it is free from closed timelike curves.

Another chronology boundary appears for $g_{55} = 0$, that is $B=0$. An ergo-region appears at $F=0$ due to the rotation in the plane ($t$, $x_5$) with radius
\begin{equation}\label{eq:5d_ergo}
    r_F^\pm = M - 2D \pm \sqrt{M^2 + N^2 + P^2 + Q^2 - D^2}.
\end{equation}
The branch $D_0$ always possesses real roots of (\ref{eq:5d_ergo}) as $D_0^2 \leq M^2 + N^2$.

\setcounter{equation}{0}
\section{Geodesics}
\label{sec:geodesics}

\subsection{Constants of motion}
The solution (\ref{eq:static_solution}) possesses the same Killing vectors as the Schwarzschild-NUT solution \cite{Clement:2015aka}
\begin{subequations}\label{eq:geo.killing}
    \begin{equation}
        K_{(t)} = \partial_t,
    \end{equation}
    \begin{equation}
        K_{(\varphi)} = \partial_\varphi,
    \end{equation}
    \begin{equation}
        K_{(x)} = 2N \frac{\cos\varphi}{\sin\theta} \partial_t - \sin\varphi \partial_\theta - \cos\varphi \cot\theta \partial_\varphi,
    \end{equation}
    \begin{equation}
        K_{(y)} = 2N \frac{\sin\varphi}{\sin\theta} \partial_t + \cos\varphi \partial_\theta - \sin\varphi \cot\theta \partial_\varphi.
    \end{equation}
\end{subequations}
The analysis of geodesic motion in the static case is similar to that for the Reissner-Nordstr\"{o}m-NUT metric (RN-NUT), which possesses spherical symmetry at the level of algebra. The geodesic motion for the RN-NUT metric was analyzed in \cite{Clement:2015aka, Zimmerman:1989kv}. The Killing vectors $K_{(t)}$ and $K_{(\varphi)}$ lead to the conservation laws of energy $E$ and angular momentum projection $J_z$
\begin{subequations}\label{eq:geodesics.static.constant_eJ}
    \begin{equation} \label{eq:geodesics.static.constant_e}
        E = f \left(\dot{t} + 2 N \cos\theta \dot{\varphi}\right),
    \end{equation}
    \begin{equation}\label{eq:geodesics.static.constant_J_z}
        J_z =
        \Sigma \sin^2\theta \dot{\varphi}
        - 2NE\cos\theta,
    \end{equation}
\end{subequations}
where the dot $\dot{}$ means the derivative with respect to the affine parameter $\tau$. In addition, the metric admits three Killing vectors ($K_{(x,y,\varphi)}$) corresponding to the generators of the rotation group $SO(3)$. The spatial Killing vectors allow to introduce a conserved total angular momentum vector $\vec{J}$, which can be divided into two parts -- the orbital angular momentum $\vec{L}$ and the spin angular momentum $\vec{S}$
\begin{equation} \label{eq:geodesics.static.sum_angular}
    \vec{J} = \vec{L} + \vec{S},
\end{equation}
with
\begin{equation}\label{eq:geodesics.static.angular_defs}
    J_i = - K_{(i)}^\mu \dot{x}_\mu, \quad
    \vec{L} = \Sigma \left[\hat{r} \times \dot{\hat{r}} \right], \quad
    \vec{S} = S \hat{r},
\end{equation}
where $S=-2 N E$ and $\hat{r}$ is a unit vector normal to the 2-sphere
\begin{equation*}
    \hat{r} = \left(
        \sin\theta \cos\varphi,
        \sin\theta \sin\varphi,
        \cos\theta
        \right).
\end{equation*}
The statement (\ref{eq:geodesics.static.sum_angular}) can be verified with the definitions (\ref{eq:geodesics.static.angular_defs}) and the substitution $\dot{t}$ from (\ref{eq:geodesics.static.constant_eJ}). From the orthogonality of $\vec{L}$ and $\vec{S}$, it follows
\begin{equation}\label{eq:geodesics.static.jls}
    J^2 = L^2 + S^2.
\end{equation}
As $J^2$ and $S^2$ are constants, $L^2$ is a constant as well. Squaring $\vec{L}$ from the definition (\ref{eq:geodesics.static.angular_defs}) gives the following relation
\begin{equation} \label{eq:geodesics.static.sphere_L}
    L^2 =
    \Sigma^2 \left(\dot{\theta}^2+\sin^2\theta \dot{\varphi}^2\right).
\end{equation}

\subsection{Angular and temporal motion}
Following \cite{Kagramanova:2010bk} and \cite{Clement:2015aka} let us introduce a new parameter, "Mino" time $\lambda$, instead of the affine parameter $\tau$, a new variable $\xi$, and divide the coordinate time $t$ into two components:
\begin{equation}
    d\tau = \Sigma d\lambda,\qquad
    \xi = \cos\theta,\qquad
    t(\lambda) = t_r(\lambda) + t_\theta(\lambda),
\end{equation}
such that
 \begin{equation}
        \label{eq:geodesics.static.tr_eq}
        t_r' = E \frac{\Sigma}{f},
    \end{equation}
where ${}'$ denotes derivative with respect to $\lambda$.
Then, equations (\ref{eq:geodesics.static.sphere_L}) and (\ref{eq:geodesics.static.constant_eJ}) can be resolved with respect to $t, \xi, \varphi$
\begin{subequations}
    \begin{equation}
        \label{eq:geodesics.static.xi_eq}
        \xi'^2 = - J^2\xi^2 +2S J_z \xi + L^2 - J_z^2,
    \end{equation}
    \begin{equation}
        \label{eq:geodesics.static.phi_eq}
        \varphi'
        = \frac{1}{2}\left(
            \frac{J_z + S}{1 + \cos\theta}
            +
            \frac{J_z - S}{1 - \cos\theta}
        \right),
    \end{equation}
    \begin{equation}
        \label{eq:geodesics.static.ttheta_eq}
        t_\theta' =
        N \left(
            -2S
            +
            \frac{S + J_z}{1 + \cos\theta}
            +
            \frac{S - J_z}{1 - \cos\theta}
        \right),
    \end{equation}
\end{subequations}
Equations for (\ref{eq:geodesics.static.xi_eq}), (\ref{eq:geodesics.static.phi_eq}) and (\ref{eq:geodesics.static.ttheta_eq}) coincide with the equations for the RN-NUT solution in Einstein-Maxwell theory, which were solved in \cite{Clement:2015aka}:
\begin{subequations}
\begin{align}
    &  \label{eq:geodesics.static.theta_functions}
        \cos\theta = \cos\psi \cos\eta + \sin\psi \sin\eta \cos(J\lambda),
    \\
    & \label{eq:geodesics.static.phi_solution}
        \phi = \phi_0
            + \arctan\left[
                \frac{\cos\psi - \cos\eta}{1 - \cos(\psi - \eta)}
                \tan\frac{J\lambda}{2}
            \right]
            + \\\nonumber & \quad\quad
            + \arctan\left[
                \frac{\cos\psi + \cos\eta}{1 + \cos(\psi - \eta)}
                \tan\frac{J\lambda}{2}
            \right],
    \\
    & \label{eq:geodesics.static.t_theta_solution}
        \frac{t_\theta}{2N} =
            -S \lambda
            - \pi \left(
                  \text{sgn}(J_z - S)
                - \text{sgn}(J_z + S)
            \right)
            \left\lfloor
                \frac{J\lambda}{2\pi} + \frac{1}{2}
            \right\rfloor
            - \\\nonumber & \quad\quad
            - \arctan\left[
                \frac{\cos\psi - \cos\eta}{1 - \cos(\psi - \eta)}
                \tan\frac{J\lambda}{2}
            \right]
            + \arctan\left[
                \frac{\cos\psi + \cos\eta}{1 + \cos(\psi - \eta)}
                \tan\frac{J\lambda}{2}
            \right],
\end{align}
\end{subequations}
where
\begin{align*}
    & J \cos\eta = S,
    & J \sin\eta = L,\\
    & J \cos\psi = J_z,
    & J \sin\psi = J_\perp,\\
\end{align*}
with
$$J_\perp^2 = J^2 - J_z^2.$$
As discussed in \cite{Clement:2015aka}, it follows from $\vec{J}\cdot\hat{r} = S$ that all the orbits with $r'=0$ are circular, whatever the plane in which they lie.

\subsection{Radial motion}
The radial equation can be obtained from the constraint $\dot{x}_\mu \dot{x}^\mu = \varepsilon$ with $\varepsilon = 1, 0, -1$ for time-like, null and space-like geodesics. Substituting $E$, $J_z$ and $L$ instead of $\dot{t}$, $\dot{\varphi}$, $\dot{\theta}$, and rewriting the equation for Mino time, one can get the radial equation
\begin{equation} \label{eq:geodesics.static.radial_mino}
   r'^2 =  E^2 \Sigma^2 - \Delta \left( \varepsilon \Sigma +L^2 \right).
\end{equation}

\textbf{Scattering on the string inside the chronology boundary.}
We will show that null and time-like geodesics can scatter near the Misner string inside the chronology boundary. For large $r$, the radial equation is $\dot{r}^2 = E^2 - \varepsilon$, imposing the only condition $E^2 \geq \varepsilon$. Now, consider the equation (\ref{eq:geodesics.static.xi_eq}) for $\xi'$ with $\xi = \pm (1 - \epsilon)$ and $J_z = \pm (2NE + \mathcal{O}(\epsilon))$   with infinitesimal $\epsilon > 0$. Then, the equation takes the form
\begin{equation}
\xi'^2 = 2L^2\epsilon + \mathcal{O}(\epsilon^2),
\end{equation}
which gives real $\xi$ for any $L$. Therefore, the geodesic curve, moving from spatial infinity, can cross the string vicinity at least when $r$ is large enough.

\textbf{Null geodesics.}
Rewriting the radial equation (\ref{eq:geodesics.static.radial_mino}) in terms
of the affine parameter $\tau$ in the form
 \be
\dot{r}^2 + U_{\text{eff}}(r) = E^2,
 \ee
the effective potential for null geodesics is
\begin{equation}\label{eq:geodesics.effective_null}
    U_{\text{eff}} = L^2 \frac{\Delta }{ A B }.
\end{equation}
It diverges at the surface of naked singularities and becomes zero at the surface of regular black holes. If the solution is a singular black hole, the behaviour of $U_{\text{eff}}$ can be either bounded or diverging, depending on the root  multiplicity  of the numerator and denominator. Another interesting question is the existence of stable or unstable photon circular orbits. The equation of circular orbits $dU_{\text{eff}}/dr = 0$ is difficult to solve analytically. Picking up a large number of different solutions with random parameters, one can find that usually regular black holes has one unstable circular orbit, while naked singularities have neither stable nor unstable circular orbits. For the case of singular black holes with $D_{\pm1}$ or $D_0$ with $e=2\mu$, there are no circular orbits. At the same time singular black holes with $D_0$ and $e=\mu$ have a maximum of the effective potential at the surface of the horizon.

Though one can conjecture these rules are general, there is a set of fine tuned counterexamples representing naked singularities, for example ${e/\mu = 1.5,}\,{\beta = \pi/2,}\,{\alpha = 3\pi/2 + \epsilon}$, where $0.099665\lesssim \epsilon \lesssim 0.123340$. Such solutions possesses both stable and unstable circular null orbits (fig. \ref{fig:eff_potential_1}). At the lower bound  $\epsilon \approx 0.099665$,   the solution becomes an extremal black hole, so the stable orbit lies on the horizon. At the upper bound  $\epsilon \approx 0.123340$ the minimum and maximum of the effective potential disappear. Generally, such solutions are small deviations (in the charge space) from extremal black holes with a singularity situated close enough to the horizon from the inner side. The existence of stable circular photon orbits indicates at the possibility to accumulate the energy of the electromagnetic field, leading to the solution instability.
\begin{figure}

    \subfloat[][] {
        \includegraphics[width=0.45\textwidth]{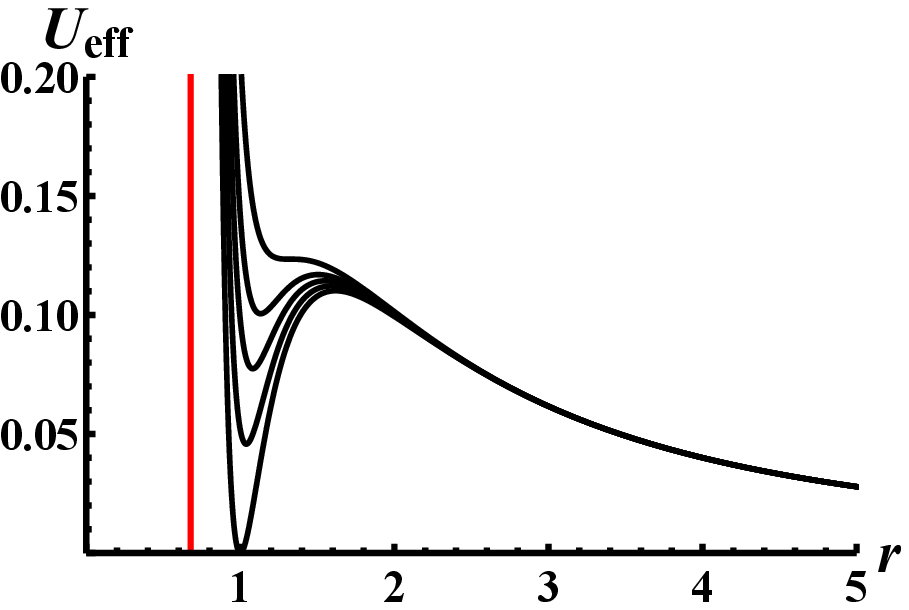}
        \label{fig:eff_potential_1}
    }
    \subfloat[][] {
        \includegraphics[width=0.45\textwidth]{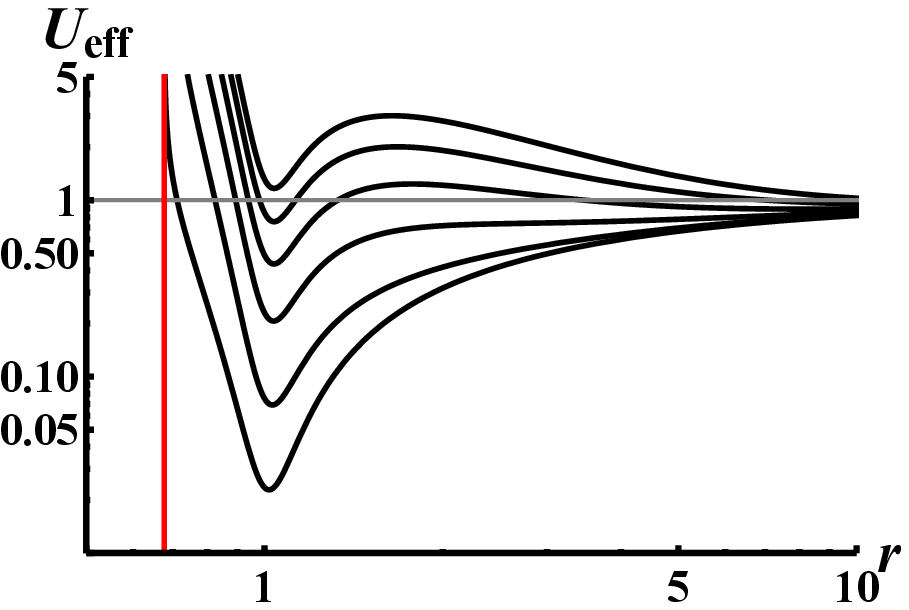}
        \label{fig:eff_potential_2}
    }

    \caption{
        Effective potential $U_{\text{eff}}$ for solutions ${\mu=1,}\,{e = 1.5,}\,{\beta = \pi/2,}\,{\alpha = 3\pi/2 + \epsilon}$. Left: $U_\text{eff}$ of null geodesics ($L^2 = 1$) for $\epsilon = 0.099665,\,0.105,\,0.11,\,0.115,\,0.123340$ (from lower to upper curves). Right: $U_\text{eff}$ of timelike geodesics with $L$ from 0 to 5 for $\epsilon=0.105$; the scale is logarithmic for better visualisation. The red line represents the naked singularity.
    }
    \label{fig:eff_potential}
\end{figure}

The effective potential (\ref{eq:geodesics.effective_null}) represents a 4th degree polynomial. Integration of this equation leads to the following solution
\begin{equation}\label{geodesics.r_eq}
    \int_{r_0}^{r} \frac{dr}{\sqrt{P_4(r)}} =
    \kappa(\lambda - \lambda_0),
\end{equation}
where $\kappa \equiv \pm \left|E\right|$, $P_4(r) =\Sigma^2 - L^2\Delta/E^2  = (r - r_1)(r - r_2)(r - r_3)(r - r_4)$ is a fourth degree polynomial, factored into its roots. The integral can be evaluated
\begin{align}\label{geodesics.r_int}
    \int_{r_0}^{r} \frac{dr}{\sqrt{P_4(r)}} \equiv I(r) - I(r_0)
    =
    \frac{ 2 }{\sqrt{(r_1 - r_3)(r_2 - r_4)}}
    \left.
    \text{F}\left(
        \arcsin\left(m(r)^{-1/2}\right)
        \Big|
        m(r_3)
    \right)
    \right|^r_{r_0},
\end{align}
\begin{equation*}
    m(r) =  \frac{(r_2-r)(r_1-r_4)}{(r_1-r)(r_2-r_4)},
\end{equation*}
where $\text{F}(\varphi|m)$ is the elliptic integral of the first kind, and the roots $r_i$ can be permuted with any order. If the chosen constant $r_0$ is placed between two real roots $r_i$ and $r_j$ (or infinity), the solution will describe the motion inside the region $r\in[r_i, r_j]$. For simplicity we will choose $r_0 = r_1$, so $I(r_0) = 0$. The function $r(\lambda)$ can be expressed from (\ref{geodesics.r_eq}) and (\ref{geodesics.r_int})
\begin{equation}\label{geodesics.r_func}
    r(\lambda) =
    \frac{
        r_1(r_4-r_2) +
        r_2 (r_1-r_4) R(\lambda)
    }{
        r_4 - r_2 + (r_1-r_4) R(\lambda)
    },\\
\end{equation}
\begin{equation*}
    R(\lambda) = \text{sn}\left(
            \frac{\kappa}{2} (\lambda - \lambda_0) \sqrt{(r_1-r_3)(r_2-r_4)}
            \Big|
            m(r_3)
        \right)^2,
\end{equation*}
where $\text{sn}(u|m)$ is the Jacobi elliptic function. Two examples of null geodesics, scattered on the black hole are shown in fig. \ref{fig:null_geodesics}. The first example (figs. \ref{fig:null_geodesics_11}, \ref{fig:null_geodesics_12}) demonstrates the family of geodesic curves crossing the chronology boundary, providing a proof of their existence. When the turning point is close enough to the potential minimum, the geodesic curve has enough time to make any number of revolutions around the horizon (fig. \ref{fig:null_geodesics_21}, \ref{fig:null_geodesics_22}).

\begin{figure}

    \subfloat[][] {
        \includegraphics[height=0.18\textheight]{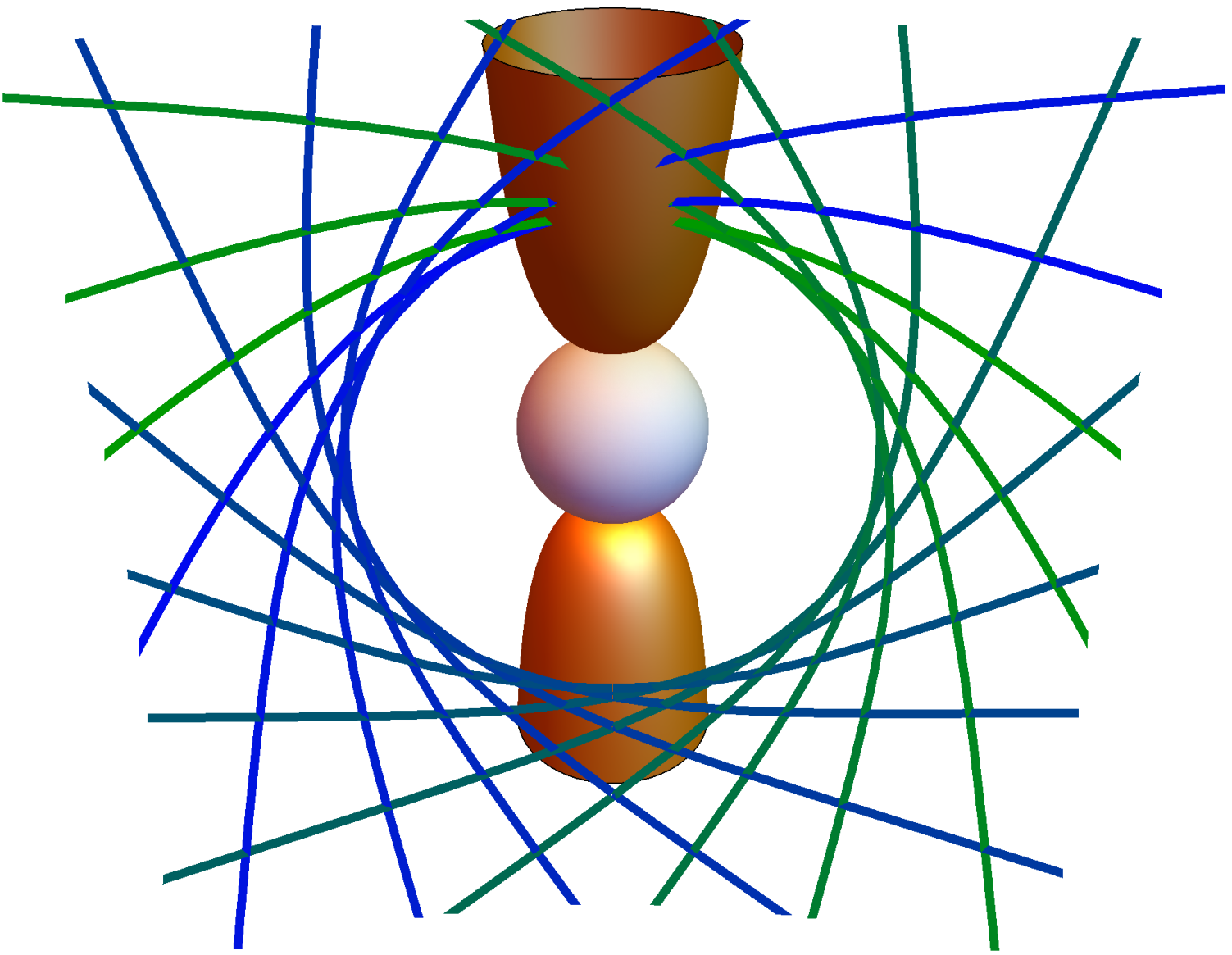}
        \label{fig:null_geodesics_11}
    }
    \subfloat[][] {
        \includegraphics[height=0.18\textheight]{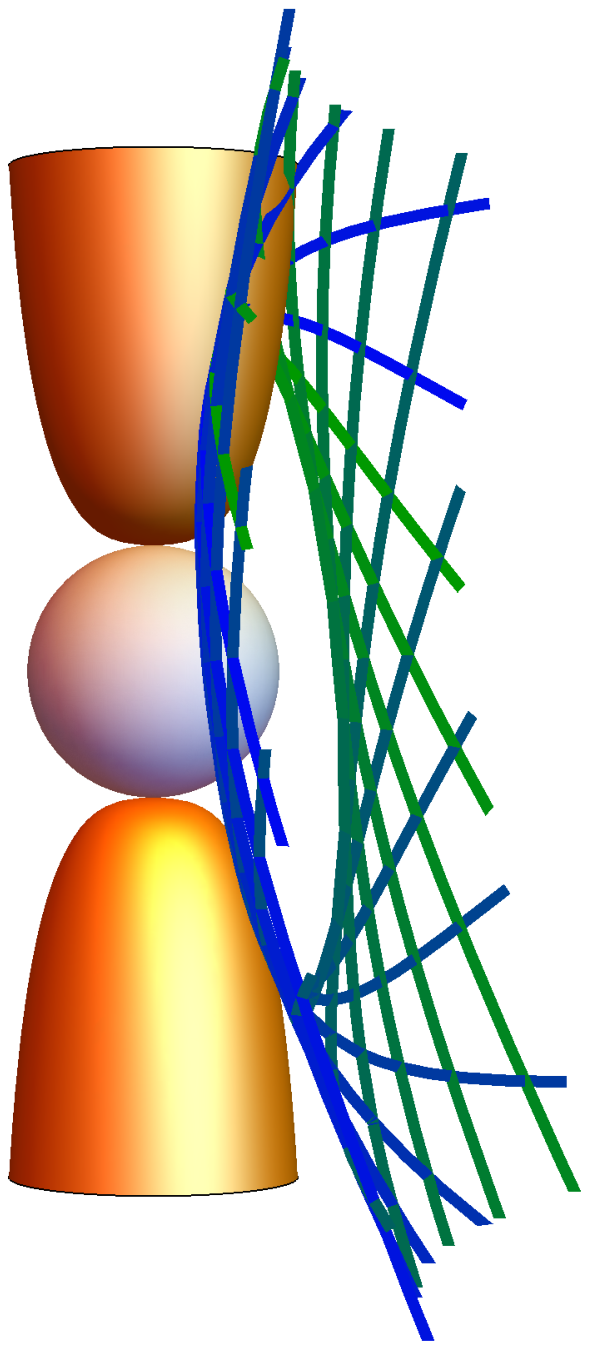}
        \label{fig:null_geodesics_12}
    }
    \hfill
    \subfloat[][] {
        \includegraphics[height=0.18\textheight]{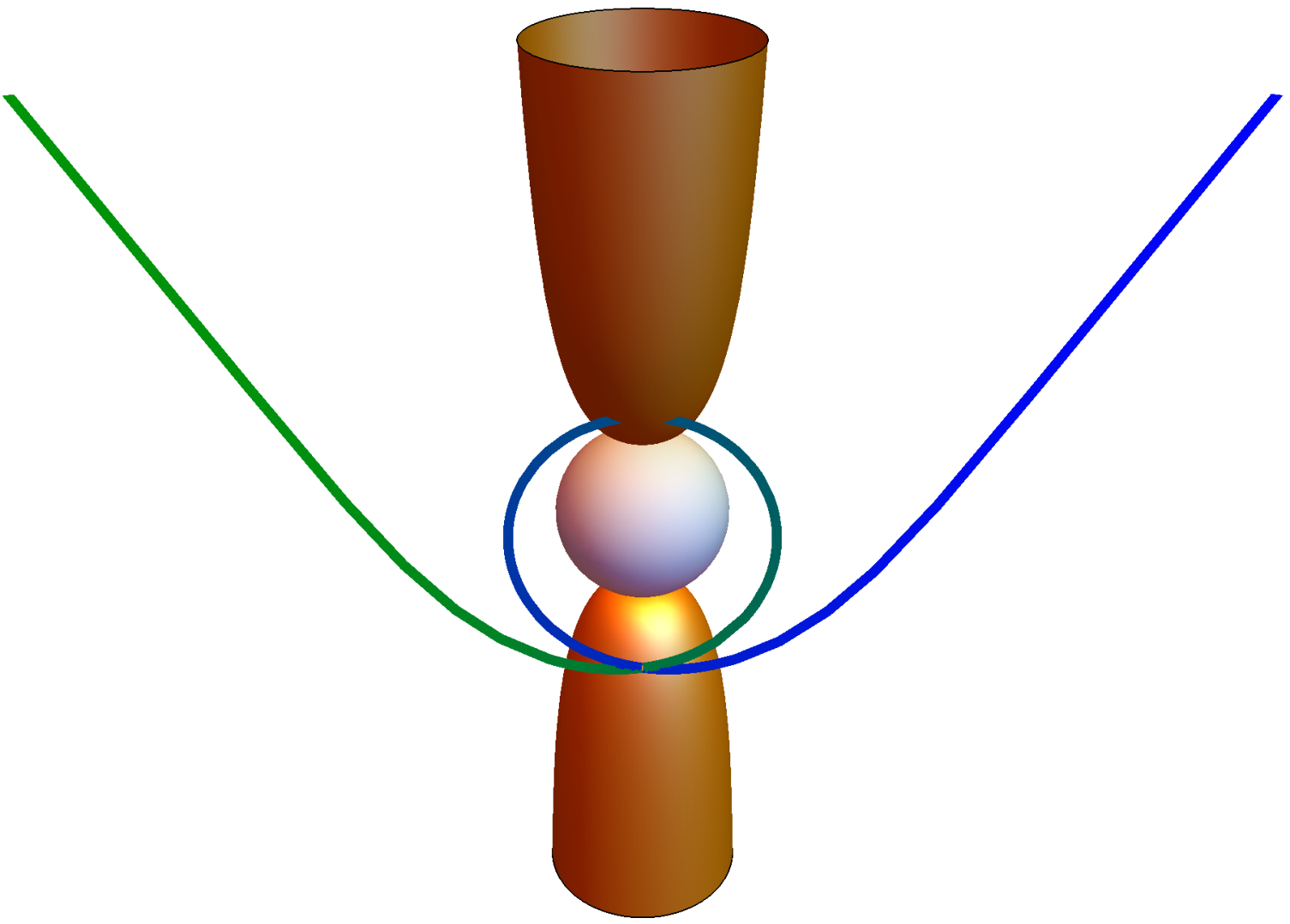}
        \label{fig:null_geodesics_21}
    }
    \subfloat[][] {
        \includegraphics[height=0.18\textheight]{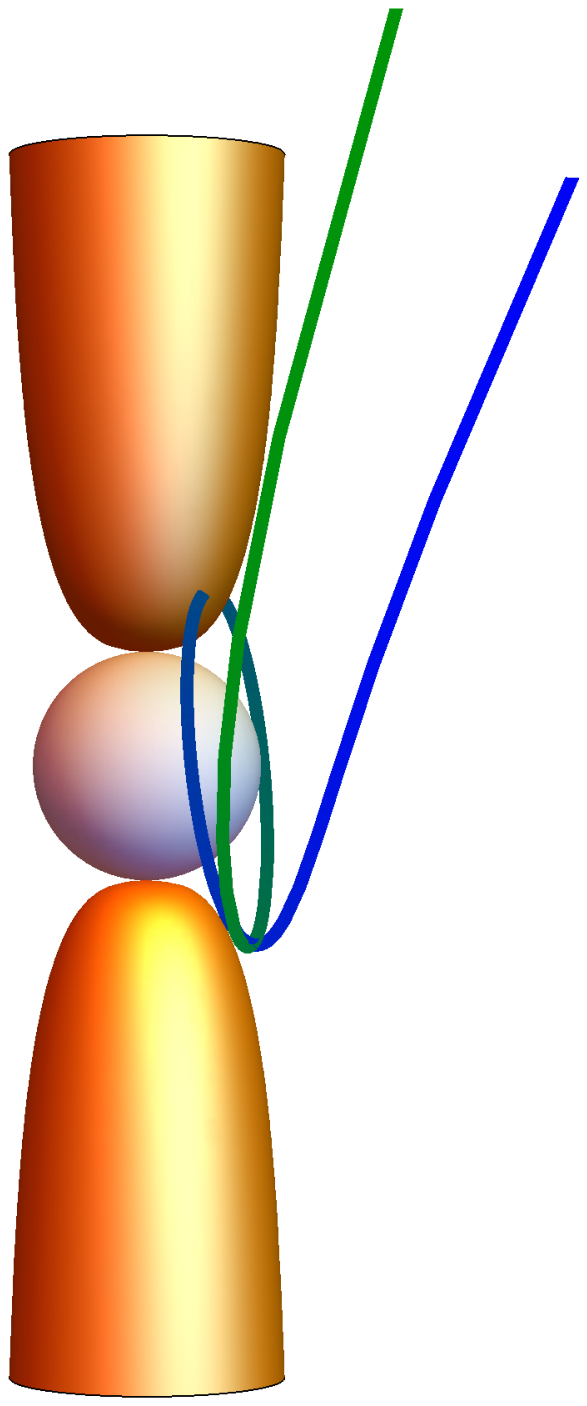}
        \label{fig:null_geodesics_22}
    }
    \caption{
        Null geodesics in the space-time with parameters $P = -0.7, Q = 1.1, D = 0.1, N = -5, M = 2.187$. The spheres are horizons and red surfaces are chronology boundaries.
        Fig. \ref{fig:null_geodesics_11}, \ref{fig:null_geodesics_12}: a family of geodesics, scattering on the Misner string ($E = 0.2,\,J = 5.5,\,J_z = 0.5$).
        Fig. \ref{fig:null_geodesics_21}, \ref{fig:null_geodesics_22}: geodesic curve with a turning point which is close to the minimum of $U_\text{eff}$, $E = 0.24148,\,J = 5.5,\,J_z = 0.5$.
    }
    \label{fig:null_geodesics}
\end{figure}

\textbf{Time-like geodesics.}
The effective potential for time-like geodesics is
\begin{equation}
    U_\text{eff} = \frac{\Delta}{\Sigma} + L^2\frac{\Delta}{\Sigma^2}.
\end{equation}
Solutions with $M \neq 0$ have one stable circular timelike orbit with $r \approx L^2 \sqrt{M^2+N^2} / M$ for large $L$. Regular solutions additionally possess an unstable circular timelike orbit near the horizon if $L$ is large enough. Usually, there are no other circular orbits, but this   is not a general rule. For the same special class of naked singularities we have considered for null geodesics, there is a stable circular timelike orbit near the singularity for any $L$ (fig. \ref{fig:eff_potential_2}).

\subsection{Closed null and timelike geodesics}
For a geodesic curve to be a closed curve, all the coordinates should again take the same values after a finite lapse of $\tau$ or $\lambda$. The Mino period of the angular functions $\theta$ and $\varphi$ is $\Delta\lambda = 2\pi/J$. Considering $J,E>0$, one can find the inequality on the corresponding $\Delta t_\theta$ from (\ref{eq:geodesics.static.t_theta_solution})
\begin{equation}\label{eq:geodesics.static.ctc_1}
    \Delta t_\theta \geq - \frac{2\pi |S|}{JE}\left(J - |S|\right)
\end{equation}
Also, using (\ref{eq:geodesics.static.tr_eq}) with $f=\Delta/\Sigma$, and the condition $r'^2 \geq 0$ in (\ref{eq:geodesics.static.radial_mino}), with $\varepsilon$ non-negative for non-tachyonic matter,
\begin{equation}\label{eq:geodesics.static.ctc_2}
    \frac{dt_r}{d\lambda} = E \frac{\Sigma}{f}
    \geq
    \frac{1}{E}\left( \Sigma\varepsilon + L^2 \right)
    \geq
    \frac{L^2}{E}.
\end{equation}
Combining all together we get
\begin{equation}\label{eq:geodesics.static.ctc_4}
    \Delta t_\theta + \Delta t_r \geq
    \frac{2\pi}{E} \left( J - |S| \right)
\end{equation}
As $J \geq \left| S \right|$ from the definition (\ref{eq:geodesics.static.jls}), this inequality ensures that $\Delta t \geq 0$ and can be saturated only for null geodesics with $J = |S|$ (i.e. $L=0$). But in this special case $\Delta t_\theta = 0$ and $\Delta t_r > 0$, so that these null geodesics cannot be closed.

\subsection{Geodesics in 5D}

In this subsection, we will denote the fifth coordinate as $x_5\equiv\chi$. Because of the existence of a non-vanishing dilaton field, leading to a non-constant $g_{\chi\chi}$, and the possibility of a non-vanishing constant momentum $p_\chi$ conjugate to the cyclic coordinate $\chi$ (which would be associated in four dimensions with the electric charge of a test particle), five-dimensional geodesic motion cannot be simply uplifted from that in four dimensions (see Appendix \ref{sec:geodesic_reduction}), but must be analyzed separately. The Killing vectors of the five-dimensional metric are
\begin{subequations}\label{eq:geo.5d_killing}
    \begin{equation}
        K_{(\chi)} = \partial_\chi,
    \end{equation}
    \begin{equation}
        K_{(t)} = \partial_t,
    \end{equation}
    \begin{equation}
        K_{(\varphi)} = \partial_\varphi,
    \end{equation}
    \begin{equation}
        K_{(x)} = \frac{\cos\varphi}{\sin\theta} ( 2P \partial_\chi +  2N \partial_t) - \sin\varphi \partial_\theta - \cos\varphi \cot\theta \partial_\varphi,
    \end{equation}
    \begin{equation}
        K_{(y)} = \frac{\sin\varphi}{\sin\theta} ( 2P \partial_\chi +  2N \partial_t) + \cos\varphi \partial_\theta - \sin\varphi \cot\theta \partial_\varphi.
    \end{equation}
\end{subequations}

Proceeding in the same way as we have done for the four-dimensional case with another affine parameter $\tau$ and another Mino time $\lambda$ such that $d\tau = A d\lambda$, one can get an analogous equation $\vec{J} = \vec{L} + \vec{S}$ with
\begin{equation}
    \vec{L} = A \left[\hat{r} \times \dot{\hat{r}} \right], \quad
    \vec{S} = S \hat{r}, \quad S = 2(Pp_\chi - N E),
\end{equation}
where $E$ is the constant momentum canonically conjugate to the cyclic coordinate $t$, and the same equations (\ref{eq:geodesics.static.xi_eq}), (\ref{eq:geodesics.static.phi_eq}), (\ref{eq:geodesics.static.ttheta_eq}) up to the redefinition of $S$ and equations
\begin{subequations}
    \begin{equation}
        t_r' = \frac{A}{\Delta }(B E+2 C p_\chi),
    \end{equation}
    \begin{equation}
        \chi_r' = \frac{A}{\Delta }(F p_\chi-2 C E),
    \end{equation}
    \begin{equation}
        \chi_\theta' = \frac{P}{N} t_\theta',
    \end{equation}
\end{subequations}
where the coordinate $\chi$ has been split into two parts, $\chi = \chi_r + \chi_\theta$. The solutions for the functions $\varphi$, $\theta$ and $t_\theta$ are again given by (\ref{eq:geodesics.static.theta_functions}), (\ref{eq:geodesics.static.phi_solution}), (\ref{eq:geodesics.static.t_theta_solution}), up to the redefinition of the constant $S$, and $\chi_\theta = P t_\theta/N + \text{const}$. The radial equation is
\begin{equation} \label{eq:radial_5d}
    r'^2 =
    A \left( B E^2 + 4 C E p_\chi - \Delta \varepsilon - F p_\chi^2 \right)
    -\Delta  L^2,
\end{equation}
which from $r'^2 \geq 0$ implies
\begin{equation}\label{eq:geodesics5d_constraint1}
    \frac{A}{B} \left( B E + 2C p_\chi \right)^2
     \geq \Delta (A \varepsilon +  L^2 + \frac{A^2}{B}p_\chi^2).
\end{equation}
In five dimensions the function $B$ may be non-positive in the physical domain (outside the horizon). At the same time, in the region $B<0$ the timelike coordinate is $\chi$, in which case the compactification of $\chi$ leads to a compact time-like direction. The radial equation can be solved similarly to (\ref{geodesics.r_eq}) for any $\varepsilon$. In the outer region with $A,B,\Delta>0$ the right-hand side of (\ref{eq:geodesics5d_constraint1}) is positive for all geodesics except purely radial null geodesics ($\epsilon=L=p_\chi=0$), therefore the left-hand side cannot be negative, which makes the surface $B E + 2C p_\chi = 0$ unreachable. Particularly, if $p_\chi = 0$, the geodesic cannot cross the surface $B=0$. Generally, the radial equation may be rewritten as
\begin{equation}
    r'^2 = AB \left(E - V_+\right)\left(E - V_-\right),\qquad
    V_\pm = 2p_\chi \frac{C}{B} \pm
    \sqrt{ \Delta \left(
         \frac{\varepsilon}{B}
        + \frac{A}{B^2} p_\chi^2
        + \frac{ L^2 }{A B}
    \right)}
\end{equation}
When a geodesic curve approaches the surface $B=0$, every branch of  the effective potential has form
\begin{equation}
    V_\pm \approx \frac{2p_\chi C \pm |p_\chi|\sqrt{\Delta A}}{B}.
\end{equation}
As the fraction $(\Delta A - 4C^2)/B$ is a polynomial (\ref{eq:oxidation.F}), then at $B=0$ we have the identity $\sqrt{\Delta A} = 2|C|$, and one of the branches $V_\pm$ is finite, but another one is diverging, depending on the sign of $C p_\chi$ (fig. \ref{fig:5D_eff_potential_1}).

Let us consider the upper potential branch $V_+$ in the region $A,B,\Delta > 0$. If the sign of $C p_\chi$ is negative and at some point $V_+=0$, then we can  observe  the Penrose process (fig. \ref{fig:5D_eff_potential_2}), extracting energy from the rotation in the plane $(t,\chi)$. From the point of view of a four-dimensional observer, the momentum $p_\chi$ corresponds to the electric charge of a test particle. Therefore, the Penrose process increases both the ``electric charge'' and the energy of the particle. 

\textbf{Traversability.} 5D wormholes are traversable if at least some timelike or null geodesics extend from one end $(r\to\infty)$ to the other end $(r\to-\infty)$. The condition for this is obviously that the right-hand side of the radial equation (\ref{eq:radial_5d}) is positive for all real $r$. If such geodesics exist, then there are radial geodesics among them $L=0$. As $A$ and $\Delta$ are positive for wormholes and $\varepsilon$ is non-negative, a necessary condition is therefore
 \be\label{eq:radial_5d_condition_1}
y(r) \equiv E^2 B(r) + 4Ep_\chi C(r) - p_\chi^2 F(r) > 0.
 \ee

Assuming $p_\chi\neq0$ (it is clear that geodesics with $p_\chi=0$ will turn back at a zero of $B(r)$, which always exists due to the absence of wormholes in 4D), the condition (\ref{eq:radial_5d_condition_1}) can be rewritten
\be
y(r) = p_\chi^2 B(r)(x-x_+(r))(x-x_-(r)),
 \ee
with $x=E/p_\chi$, and 
 \be
x_\pm = \frac{-2C\pm\sqrt{A\Delta}}B.
 \ee
The allowed range ($y>0$) is thus: 
 \ba
{\rm For}\; B(r) > 0 & (x_+>x_-): & x > x_+(r) \quad {\rm or} \quad x < x_-(r), \\
{\rm For}\; B(r) <  0 & (x_+<x_-): & \qquad x_+(r) < x < x_-(r).
 \ea
At infinity, $B(r)> 0$, and $x_\pm(r) \to \pm 1$, so that geodesics coming from or extending to infinity
(either wormhole end) must have $x>1$ or $x<-1$.

The function $B(r)$ has two zeroes, $r_{B-} < r_{B+}$. Assume without loss of generality $C(r_{B+}) > 0$. Then,
 \be
x_+(r_{B+}) = \frac{F(r_{B+})}{4C(r_{B+})}, \quad x_-(r) \simeq - \frac{4C(r_{B+})}{B(r)} \;\; (r\simeq r_{B+}).
 \ee
Geodesics coming from $r \to + \infty$ with $x < -1$ must be such that $x < x_-(r)$, and thus will necessarily turn back at 
some $r_{\rm min}$ before reaching $r_{B+}$. Only geodesics with $x>1$ can possibly extend further.

The function $C(r)$ has a single zero $r_C$, and the outcome depends on whether $B(r_C)$ is positive or negative.

1) \underline{$B(r_C)>0$}. Then $C(r_{B-}) > 0$, and
 \be
x_+(r_{B-}) = \frac{F(r_{B-})}{4C(r_{B-})}, \quad x_-(r) \simeq - \frac{4C(r_{B-})}{B(r)} \;\; (r\simeq r_{B-}).
 \ee
So, in the range $r_{B-} < r < r_{B+}$, the curve $x_-(r)$ goes to $+\infty$ at both ends $r = r_{B\pm}$, and lies above the curve 
$x_+(r)$. Geodesics coming from infinity can therefore go through provided $x_{- \rm min}$ (the relative minimum of $x_-(r)$) 
is larger than 1, and can proceed to the other wormhole end if $x_{- \rm min} > x_{+ \rm max}$ (the absolute maximum of $x_+(r)$).
  
2) \underline{$B(r_C)<0$}. Then $C(r_{B-}) < 0$, and
 \be
x_+(r) \simeq \frac{4|C(r_{B-})|}{B(r)} \;\; (r\simeq r_{B-}), \quad x_-(r_{B-}) = - \frac{F(r_{B-})}{4|C(r_{B-})|}, 
 \ee-
In the range $r < r_{B+}$, the curve $x_-(r)$ goes continuously from $+\infty$ at $r = r_{B+}$ to $-1$ for $r \to -\infty$, and 
so must cross the line $x=+1$ for some finite value of $r$. All geodesics with $x>1$ will necessarily turn back at a value 
$r_{\rm min} < r_{B+}$ such that $x_-(r_{\rm min}) = x$.

\begin{figure}

    \subfloat[][] {
        \includegraphics[width=0.45\textwidth]{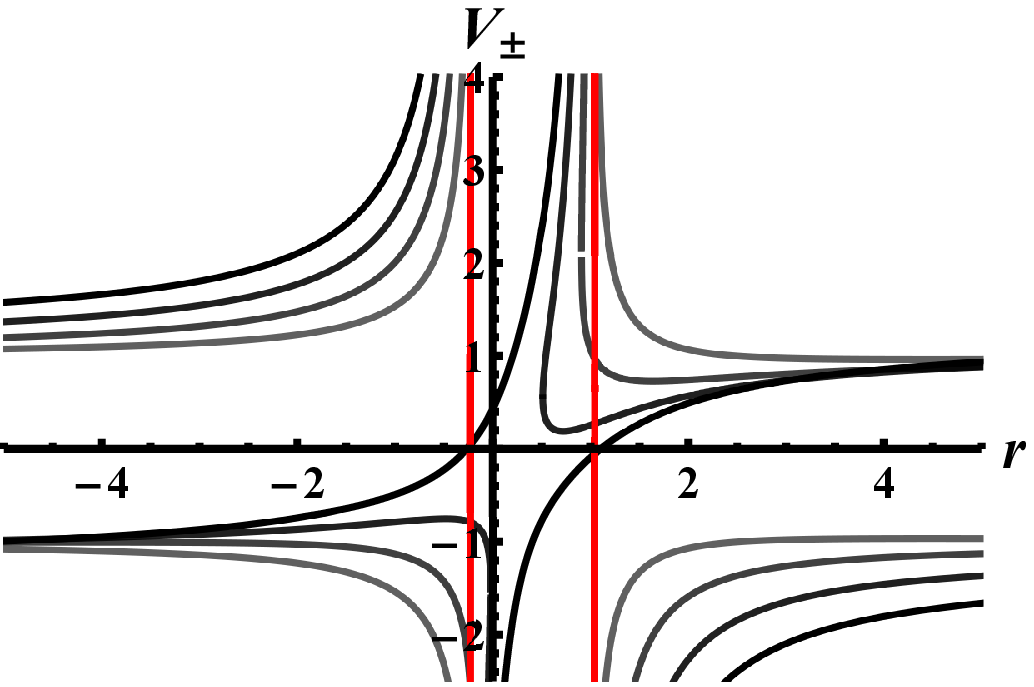}
        \label{fig:5D_eff_potential_1}
    }
    \subfloat[][] {
        \includegraphics[width=0.45\textwidth]{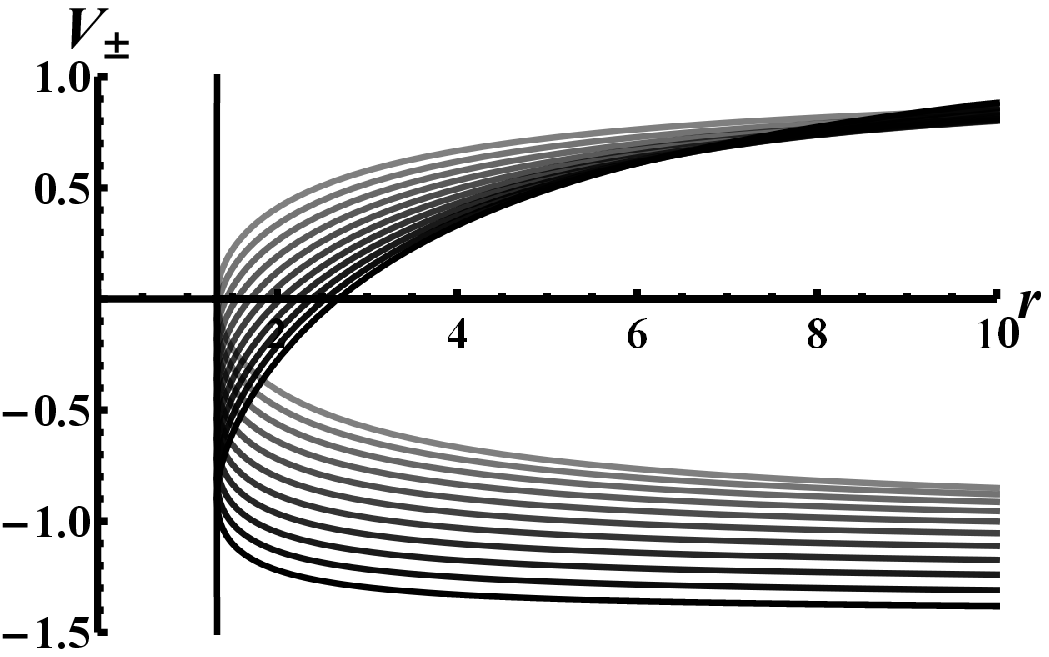}
        \label{fig:5D_eff_potential_2}
    }

    \caption{
        Left: effective potential $V_{\pm}$ of a wormhole ${\mu=1,}\,{e = 1.5,}\,{\beta = 3\pi/4,}\,{\alpha = 3\pi/4}$ for $L=0.1$ and different $p_\chi$ from $0$ to $-0.75$ with a step $0.25$. Right: effective potential {\bf $V_{\pm}$} of a black hole ${\mu=1,}\,{e = 1.5,}\,{\beta = \pi/2,}\,{\alpha = 3\pi/2}$ for $L=0.1$ and different $p_\chi$ from $0$ to $-1$ with a step $0.1$. The red line is $B=0$ and the black vertical line is a horizon. The darker curve stands for larger $|p_\chi|$.
    }
    \label{fig:5D_eff_potential}
\end{figure}

\setcounter{equation}{0}
\section{Conclusions}\label{sec:conclusions}

Building on the pioneering work of \cite{Dobiasch:1981vh}, we have constructed and analyzed the general non-rotating locally asymptotically flat solution of five-dimensional vacuum gravity with two Killing vectors (one of which is timelike), and its reduction to four dimensions as a solution of $\alpha=\sqrt{3}$ EMD.  The constructive charge-matrix approach we have followed is complementary to the solution-generating approach followed e.g. in \cite{Rasheed:1995zv} -- applying special $SL(3,R)$ group transformations to the Schwarzschild solution embedded in 5D. This leads to a wider class of dyonic solutions, possessing also a NUT charge and a dilaton charge which is not related to the other charges by the usual cubic regularity constraint. These solutions include as special cases both the FJNW solutions with singular horizon and the regular locally asymptotically flat KK black holes.

We found that the cubic constraint, which states that the $sl(3,R)$ charge matrix is degenerate, is only a necessary condition for regularity of the horizon. A second condition is the proper choice of a particular solution of the cubic constraint from three possibilities. The two other solutions of the cubic equation lead to generically singular solutions belonging to the degenerate type. Also, the condition of extremal solutions was generalized for arbitrary NUT charge.

In all cases we included an independent NUT charge, with the hope that it could perhaps convert superextremal solutions into wormholes, as in the case of the Brill solution of Einstein-Maxwell theory. But in the KK theory this turned out not to be possible: no combination of five charges can give rise to a four-dimensional non-rotating wormhole. Still, there exist five-dimensional solutions of the KK theory with the wormhole topology, but these wormholes are not geodesically traversable.

From the analysis of geodesics in the background of the obtained NUTty solutions, we found that time-like and null geodesics cannot be closed in the vicinity of the polar axes inside the chronology violating region surrounding the Misner string, and showed that time-like geodesics include a class of geodesics with circular orbits lying in arbitrary (generically non-equatorial) planes.

\appendix
\renewcommand{\theequation}{\thesection.\arabic{equation}}

\setcounter{equation}{0}
\section{Four-dimensional wormholes}
\label{sec:4d_wormholes}

Einstein-Maxwell theory admits traversable wormhole solutions with NUT \cite{Clement:2015aka}. So one could expect that at least some of the five-dimensional wormholes of Kaluza-Klein theory reduce to four-dimensional wormholes of EMD  with $\alpha^2 = 3$. However, these were not observed in the numerical investigations of Sect. IV. Here we give a rigorous proof of their non-existence.

A four-dimensional  wormhole must satisfy, in addition to the constraint (\ref{eq:sigma.charge_constraint}), the conditions that the quadratic functions $A$, $B$ and $\Delta$ be positive for all real $r$, i.e.
 \be\lb{AB}
N^2 - 2D^2 > |Q^2-P^2 + 2MD| \ge 0
 \ee
and $\delta_H^2 < 0$. Putting
 \be
e_{\pm} \equiv Q^2 \pm P^2\,,
 \ee
those conditions read
 \ba
|e_- + 2MD| &<& N^2 - 2D^2\,, \lb{AB1}\\
e_+ &>& M^2 + N^2 + 3D^2\,. \lb{Da2}
 \ea
So a strategy is to eliminate e.g. $e_+$ in terms of $e_-$ using the constraint (\ref{eq:sigma.charge_constraint}), enforce the bound (\ref{Da2}), and see
whether this is consistent with the bound (\ref{AB1}).

The constraint (\ref{eq:sigma.charge_constraint}) may be written as the quadratic
equation for $e_+$:
 \be\lb{eqe+}
(N^2-D^2)e_+^2 - 2D(Me_-+2z^2D)e_+ - [(M^2+N^2)e_-^2 +
4z^2MDe_- + 4z^2D^2] = 0\,,
 \ee
where we have put
 \be
z^2 \equiv M^2+N^2-D^2
 \ee
(positive by virtue of (\ref{AB})). The discriminant
 \be
\Delta_1 = z^2N^2[(e_- + 2MD)^2 + 4D^2(N^2-D^2)]
 \ee
is positive definite. Solving (\ref{eqe+}) for $e_+$, we obtain
 \be
e_+ = \dfrac{MD(e_- + 2MD) + 2(N^2-D^2)D^2 \pm
\sqrt{\Delta_1}}{N^2-D^2} > z^2 + 4D^2
 \ee
by virtue of (\ref{Da2}). Putting
 \be
x \equiv e_- + 2MD\,,
 \ee
this reads
 \be\lb{Da3}
MDx \pm \sqrt{\Delta_1} > (z^2 + 2D^2)(N^2-D^2)\,.
 \ee
Because
 \be
\Delta_1 - M^2D^2x^2 = (N^2-D^2)[(M^2+N^2)x^2 + 4z^2D^2N^2] >
0\,,
 \ee
Eq. (\ref{Da3}) can only be satisfied for the up sign. This can be
rewritten as
 \be\lb{Da4}
\sqrt{\Delta_1} > (M^2+N^2+D^2)(N^2-D^2) - MDx\,.
 \ee
The right-hand side is positive, whatever the sign of $x$, because
 \be
|MDx| < |MD|(N^2-2D^2) < |2MD|(N^2-D^2) < (M^2+N^2+D^2)(N^2-D^2)\,.
 \ee
Therefore, squaring (\ref{Da4}) leads to the bound
 \be\lb{Da5}
(M^2+N^2)x^2 + 2MD(z^2+2D^2)x + [4N^2D^2z^2 -
(N^2-D^2)(z^2+2D^2)^2] > 0\,.
 \ee
The corresponding discriminant is simply
 \be
\Delta_2 = N^2z^6\,.
 \ee
So (\ref{Da5}) is solved by
 \be
x > x_+ > 0 \;\; {\rm or} \;\; x < x_- < 0 \,,\qquad x_\pm =
\dfrac{-MD(z^2+2D^2) \pm Nz^3}{M^2+N^2}\,,
 \ee
where we have used the fact that $x_+x_- < 0$, because the
discriminant of the last term in (\ref{Da5}), considered as a
quadratic function of $z^2$ is (proportional to)
 \be
\Delta_3 = - 4N^2(N^2-2D^2) < 0\,.
 \ee

Conversely, (\ref{Da5}) cannot be statisfied if
 \be\lb{nogo}
x_- \le x \le x_+\,.
 \ee
As $|x| < N^2-2D^2$ from (\ref{AB1}), a sufficient condition for
(\ref{nogo}) is
 \be
\pm x_\pm \ge N^2-2D^2\,,
 \ee
which can be rewritten as
 \be
-x_+x_- \ge  (N^2-2D^2)^2\,,
 \ee
or in full,
 \be
(N^2-D^2)(M^2+N^2+D^2)^2 - 4N^2D^2(M^2+N^2-D^2) -
(M^2+N^2)(N^2-2D^2)^2 \ge 0\,.
 \ee
This can be expressed as a bound for the quadratic polynomial in
$M^2$:
 \be\lb{nogo1}
(N^2-D^2)M^4 + (N^4-6D^4)M^2 + D^2(N^4 - N^2D^2 - D^4) \ge 0\,.
 \ee
The left-hand side of (\ref{nogo1}) is positive, unless the
corresponding discriminant
 \be
\Delta_4 = 16(y-1)(y^3-y^2-2y-2) \qquad (y \equiv N^2/2D^2 \ge 1)
 \ee
is positive, and $M^2$ lies between the resulting two real roots
$M^2_\pm(y)$, provided these are positive. The product of the two
roots is proportional to $(4y^2-2y-1)/(2y-1)$, which is positive for
$y>1$, so the two roots could be positive if their sum, proportional
to $-(2y^2-3)/(2y-1)$, was positive, i.e. for $1 < y < \sqrt{3/2}$.
However $\Delta_4$ is negative in that range (complex roots), so
that (\ref{nogo1}) is identically satisfied. It follows that there
are no non-rotating stationary wormhole solutions to the reduced four-dimensional theory (EMD with $\alpha^2 = 3$).

\setcounter{equation}{0}
\section{Reduction of the geodesic equation}
\label{sec:geodesic_reduction}
The reduction of geodesics in 5D to 4D was considered in \cite{Kovacs:1984qx}. Here we will perform analogous calculations using the general metric ansatz from the Kaluza-Klein theory. Consider equation of geodesics in 5D with metric $G_{MN}$ in the form
\begin{equation}\label{eq:geor.5d_metric}
    ds^2 = G_{MN}dx^M dx^N = -\e^{4\alpha\phi/3} (d\chi + 2A_\mu dx^\mu)^2 + \e^{-2\alpha\phi/3} g_{\mu\nu}dx^\mu dx^\nu,
\end{equation}
\begin{subequations}\label{eq:geor.gh_1}
    \begin{equation}\label{eq:geor.geodesics5d_1}
        \frac{d}{d\lambda} \left( h^{-1} G_{MA} u^M \right) - \frac{1}{2} h^{-1} G_{MN,A}u^M u^N = 0,
    \end{equation}
    \begin{equation}\label{eq:geor.h_1}
        h^2 m^2 = G_{MN} u^M u^N,
    \end{equation}
\end{subequations}
where $g_{\mu\nu}$, $A_\mu$ and $\phi$ depends on $x^\mu$ only, $m$ is the mass of the particle, $\lambda$ is a parameter (not necessarily affine), $u^M$ is the 5-velocity, uppercase Latin indices belong to 5D and Greek indices belong to 4D. From the equation (\ref{eq:geor.geodesics5d_1}) with $A=\chi$ one can find
\begin{equation}\label{eq:geor.p}
    u^\chi = \e^{-4\alpha\phi/3} h p_\chi -2A_\mu u^\mu,
\end{equation}
where $p_\chi$ is the conjugate momentum along the coordinate $\chi$. Rewriting (\ref{eq:geor.gh_1}) in terms of 4D quantities, substituting (\ref{eq:geor.p}) and choosing a paramerization such that $h=\e^{-2\alpha\phi/3}$, one can get
\begin{subequations}\label{eq:geor.gh_2}
    \begin{equation}\label{eq:geor.geodesics5d_2}
        \frac{d}{d\lambda} \left( g_{\mu\rho} u^\mu \right) - \frac{1}{2} g_{\mu\nu,\rho} u^\mu u^\nu =
        2p_\chi F_{\rho\mu}u^\mu - \frac{\alpha}{3} \left(
            g_{\mu\nu} u^\mu u^\nu + 2p_\chi^2 \e^{-2\alpha\phi}
        \right)\partial_\rho \phi,
    \end{equation}
    \begin{equation}\label{eq:geor.h_2}
        g_{\mu\nu}u^\mu u^\nu = m^2_\text{eff}(\phi),
    \end{equation}
\end{subequations}
where we introduced the local effective mass $m^2_\text{eff}(\phi) = m^2 \e^{-2\alpha\phi/3} + p_\chi^2 \e^{-2\alpha\phi}$, and $p_\chi$ plays the role of the effective electric charge.
Substituting (\ref{eq:geor.h_2}) in (\ref{eq:geor.geodesics5d_2}), we get the final expression
\begin{equation}\label{eq:geor.geodesics_final}
    \frac{d}{d\lambda} \left( g_{\mu\rho} u^\mu \right) - \frac{1}{2} g_{\mu\nu,\rho} u^\mu u^\nu =
    2p_\chi F_{\rho\mu}u^\mu
    -
    \frac{1}{2} \partial_\rho m^2_{\text{eff}}.
\end{equation}
Equation (\ref{eq:geor.geodesics_final}) represents the usual 4D geodesic equation of a particle with mass $m_\text{eff}$, electric charge $p_\chi$ (the coefficient 2 in the equation appears due to our definition of metric) and some additional force from the gradient of the scalar field. The particle is 4-null, and does not see the existence of the fifth dimension, if and only if $m = p_\chi = 0$ .


\begin{thebibliography}{9}

\bibitem{leut}
  Leutwyler, H. (1960).
  Arch. Sci., {\bf 13}, 549.

\bibitem{Chodos:1980df}
  Chodos, A., and Detweiler, S.L. (1982).
  Gen. Rel. Grav., {\bf 14}, 879.

\bibitem{Dobiasch:1981vh}
  Dobiasch, P., and Maison, D. (1982).
  Gen. Rel. Grav., {\bf 14}, 231.

\bibitem{Sorkin:1983ns}
  Sorkin, R.~D. (1983).
  Phys. Rev. Lett., {\bf 51}, 87.

\bibitem{Gross:1983hb}
  Gross, D.~J., and Perry, M.~J. (1983).
  Nucl. Phys. B, {\bf 226}, 29.

\bibitem{Clement:1985gm}
  Cl\'ement, G. (1986).
  Gen. Rel. Grav., {\bf 18}, 861.

\bibitem{Clement:1986bt}
  Cl\'ement, G. (1986).
  Phys. Lett. A, {\bf 118}, 11.

\bibitem{Gibbons:1985ac}
  Gibbons, G.~W., and Wiltshire, D.~L. (1986).
  Annals Phys., {\bf 167}, 201.
  [Erratum: (1987). Annals Phys., {\bf 176}, 393].

\bibitem{Rasheed:1995zv}
  Rasheed, D. (1995).
  Nucl. Phys. B, {\bf 454}, 379.
  [arXiv: hep-th/9505038].


\bibitem{neuge}
  Neugebauer, G. (1969).
  Habilitationsschrift, FSU Jena.

\bibitem{Maison:1979kx}
  Maison, D. (1979).
  Gen. Rel. Grav., {\bf 10}, 717.

\bibitem{BeRu80}
  Belinski, V., and Ruffini, R. (1980).
  Phys. Lett. B, {\bf 89},195.

\bibitem{Clement:1986dn}
  Cl\'ement, G. (1986).
  Gen. Rel. Grav., {\bf 18}, 137.

\bibitem{Frolov:1987rj}
  Frolov, V.~P., Zelnikov, A.~I., and Bleyer, U. (1987).
  Annalen Phys., {\bf 44}, 371.

\bibitem{matos}
  Matos, T. (1994).
  Journ. Math. Phys., {\bf 35}, 1302.
  [arXiv: gr-qc/9401009].

\bibitem{Poletti:1995yq}
  Poletti, S.~J., Twamley, J., and Wiltshire, D.~L. (1995).
  Class. Quant. Grav., {\bf 12}, 1753.
  [arXiv: hep-th/9502054].
  [Erratum: (1995). Class. Quant. Grav., {\bf 12}, 2355].

\bibitem{Aliev:2008wv}
  Aliev, A. N., Cebeci, H., and Dereli, T. (2008).
  Phys. Rev. D, {\bf 77} 124022.
  [arXiv: 0803.2518 [hep-th]].

\bibitem{Gibbons:1982ih}
  Gibbons, G.~W. (1982).
  Nucl. Phys. B, {\bf 207}, 337.

\bibitem{Gibbons:1984hy}
  Gibbons, G. W. and Perry, M. J. (1984).
  Nucl. Phys. B, {\bf 248}, 629.

\bibitem{Gibbons:1987ps}
  Gibbons, G.~W., and Maeda, K.~I. (1988).
  Nucl. Phys. B, {\bf 298}, 741.

\bibitem{Breitenlohner:1987dg}
  Breitenlohner, P., Maison, D., and Gibbons, G.~W. (1988).
  Commun. Math. Phys., {\bf 120}, 295.

  \bibitem{Garfinkle:1990qj}
  Garfinkle, D., Horowitz, G.~T., and Strominger, A. (1991).
  Phys. Rev. D, {\bf 43}, 3140
  [Erratum: (1992). Phys. Rev. D, {\bf 45}, 3888].

\bibitem{Ortin:2015hya}
  Ort\'{i}n, T. (2015).
  { \em Gravity and Strings}, 2nd edition. CUP.

\bibitem{Cvetic:1994hv}
  Cvetic, M., and Youm, D. (1995).
  Nucl. Phys. B, {\bf 438}, 182.
  [arXiv: hep-th/9409119].
  [Addendum: (1995). Nucl. Phys. B, {\bf 449}, 146].

\bibitem{Giusto:2007fx}
  Giusto, S., and Saxena, A. (2007).
  Class. Quant. Grav., {\bf 24}, 4269.
  [arXiv: 0705.4484 [hep-th]].

\bibitem{Niarchos:2008jc}
  Niarchos, V. (2008).
  Mod. Phys. Lett. A, {\bf 23}, 2625
  [arXiv: 0808.2776 [hep-th]].

\bibitem{Tomizawa:2008rh}
  Tomizawa, S., and Ishibashi, A. (2008).
  Class. Quant. Grav., {\bf 25}, 245007.
  [arXiv: 0807.1564 [hep-th]].

\bibitem{Horowitz:2011cq}
  Horowitz, G.~T., and Wiseman, T. (2011).
  [arXiv: 1107.5563 [gr-qc]].

\bibitem{Herdeiro:2015waa}
  Herdeiro, C.~A.~R., and Radu, E. (2015).
  Int. J. Mod. Phys. D, {\bf 24}, 9, 1542014.
  [arXiv: 1504.08209 [gr-qc]].

\bibitem{Mejias:2019aio}
  Mejías, R.~R. (2020.
  Phys. Rev. D, {\bf 101}, 124030
  [arXiv: 1907.10681 [hep-th]].

\bibitem{Astefanesei:2019pfq}
  Astefanesei, D., Herdeiro, C., Pombo, A., and Radu, E. (2019).
  JHEP, {\bf 1910}, 078.
  [arXiv: 1905.08304 [hep-th]].

\bibitem{Grunau:2019bsd}
  Grunau, S., and Kruse, M. (2020).
  Phys. Rev. D, {\bf 101}, 2, 024051.
  [arXiv: 1910.09835 [gr-qc]].


\bibitem{Azeyanagi:2008kb}
  Azeyanagi, T., Ogawa, N., and Terashima, S. (2009).
  JHEP, {\bf 0904}, 061.
  [arXiv: 0811.4177 [hep-th]].

\bibitem{Goldstein:2009cv}
  Goldstein, K., Kachru, S., Prakash, S., and Trivedi, S.~P. (2010).
  JHEP, {\bf 1008}, 078.
  [arXiv: 0911.3586 [hep-th]].


\bibitem{Hirschmann:2017psw}
  Hirschmann, E.~W., Lehner, L., Liebling, S.~L., and Palenzuela, C. (2018).
  Phys. Rev. D, {\bf 97}, 6, 064032.
  [arXiv: 1706.09875 [gr-qc]].

\bibitem{Jai-akson:2017ldo}
  Jai-akson, P., Chatrabhuti, A., Evnin, O., and Lehner, L. (2017).
  Phys. Rev. D, {\bf 96}, 4, 044031.
  [arXiv: 1706.06519 [gr-qc]].

\bibitem{McCarthy:2018zze}
  McCarthy, F., Kubizňák, D., and Mann, R.~B. (2018).
  Phys. Rev. D, {\bf 97}, 10, 104025.
  [arXiv: 1803.01862 [gr-qc]].

\bibitem{Jusufi:2018gnz}
  Jusufi, K., Banerjee, A., Gyulchev, G., and Amir, M. (2019).
  Eur. Phys. J. C, {\bf 79}, 1, 28.
  [arXiv: 1808.02751 [gr-qc]].

\bibitem{Galtsov:2018xuc}
  Gal'tsov, D., and Zhidkova, S. (2019).
  Phys. Lett. B, {\bf 790}, 453.
  [arXiv: 1808.00492 [hep-th]].

\bibitem{Domenech:2019syf}
  Domènech, G., Naruko, A., Sasaki, M., and Wetterich, C. (2020).
  Int. J. Mod. Phys. D, {\bf 29}, 3, 2050026.
  [arXiv: 1912.02845 [gr-qc]].

\bibitem{BenAchour:2019fdf}
  Ben Achour, J., Liu, H., and Mukohyama, S. (2020).
  JCAP, {\bf 2002}, 023.
  [arXiv: 1910.11017 [gr-qc]].

\bibitem{Clement:2015cxa}
  Cl\'ement, G., Gal'tsov, D., and Guenouche, M. (2015).
  Phys. Lett. B , {\bf 750}, 591.
  [arXiv: 1508.07622[hep-th]].

\bibitem{Clement:2015aka}
  Cl\'ement, G., Gal'tsov, D., and Guenouche, M. (2016).
  Phys. Rev. D, {\bf 93} 2, 024048.
  [arXiv: 1509.07854 [hep-th]].

\bibitem{Galtsov:1995mb}
  Galtsov, D.~V., Garcia, A.~A., and Kechkin, O.~V. (1995).
  Class. Quant. Grav., {\bf 12}, 2887.
  [arXiv: hep-th/9504155].

\bibitem{Galtsov:2014wxl}
  Gal'tsov, D., Khramtsov, M., and Orlov, D. (2015)
  Phys. Lett. B, {\bf 743}, 87.
  [arXiv: 1412.7709 [hep-th]].

  \bibitem{Neugebauer:1969wr}
  Neugebauer, G., and Kramer, D. (1969).
  Annalen Phys., {\bf 24}, 62.

\bibitem{Larsen:1999pp}
  Larsen, F. (2000).
  Nucl. Phys. B, {\bf 575}, 211.
  [arXiv: hep-th/9909102].


\bibitem{MasoodulAlam:1993ea}
  Masood-ul-Alam, A.~K.~M. (1993).
  Class. Quant. Grav., {\bf 10}, 2649.

\bibitem{Yazadjiev:2010bj}
  Yazadjiev, S.~S. (2010).
  Phys. Rev. D, {\bf 82}, 124050.
  [arXiv: 1009.2442 [hep-th]].

\bibitem{Chrusciel:2012jk}
  Chrusciel, P.~T., Lopes Costa, J., and Heusler, M. (2012).
  Living Rev. Rel., {\bf 15}, 7.
  [arXiv: 1205.6112 [gr-qc]].

 \bibitem{Fisher:1948yn}
  Fisher, I.~Z. (1948).
  Zh.~Eksp.~Teor.~Fiz., {\bf 18}, 636.
  [arXiv: gr-qc/9911008].

\bibitem{Bergmann:1957zza}
  Bergmann, O., and Leipnik, R. (1957).
  Phys. Rev., {\bf 107}, 1157.

\bibitem{Penney:1968zz}
  Penney, R. (1968).
  Phys. Rev., {\bf 174}, 1578.

\bibitem{Janis:1968zz}
  Janis, A.~I., Newman, E.~T., and Winicour, J. (1968).
  Phys. Rev. Lett., {\bf 20}, 878.

\bibitem{Abdolrahimi:2009dc}
  Abdolrahimi, S., and Shoom, A. A. (2010).
  Phys. Rev. D, {\bf 81}, 024035.
  [arXiv: 0911.5380 [gr-qc]]


\bibitem{Gyu}
  Gyulchev, G., Nedkova, P., Vetsov, T., and Yazadjiev, S. (2019).
  Phys. Rev. D, {\bf 100}, 024055.
  [arXiv: 1905.05273[gr-qc]].

\bibitem{Jus}
  Jusufi, K., Banerjee, A., Gyulchev, G., and Amir, M. (2019).
  Eur. Phys. J. C, {\bf 79}, 1, 28.
  [arXiv: 1808.02751[gr-qc]].

\bibitem{Israel:1972vx}
  Israel, W., and Wilson, G. A. (1972).
  J. Math. Phys., {\bf 13}, 865.

\bibitem{Horne:1992bi}
  Horne, J. H., and Horowitz, G. T. (1993).
  Nucl. Phys. B, {\bf 399}, 169.

\bibitem{Chen:2000yi}
  Chen, C.~M. (2001).
  Class. Quant. Grav., {\bf 18}, 4179.
  [arXiv: gr-qc/0009042].

\bibitem{AzregAinou:1990zp}
  Azreg-Ainou, M., and Cl\'ement, G. (1990).
  Gen. Rel. Grav., {\bf 22}, 1119.

\bibitem{AzregAinou:1999}
  Azreg-Ainou, M., Cl\'ement, G., Constantinidis, C.P., and Fabris, J.~C. (2000).
  Grav. Cosmol., {\bf 6}, 207.
  [arXiv: gr-qc/9911107 [gr-qc]].

\bibitem{Zimmerman:1989kv}
  Zimmerman, R. L., and Shahir, B. Y. (1989).
  Gen. Rel. Grav., {\bf 21}, 821.

\bibitem{Kagramanova:2010bk}
  Kagramanova, V., Kunz, J., Hackmann, E., and Lammerzahl, C. (2010).
  Phys. Rev. D, {\bf 81}, 124044.
  [arXiv: 1002.4342[gr-qc]].

\bibitem{Kovacs:1984qx}
  Kovacs, D. (1984).
  Gen. Rel. Grav., {\bf 16}, 645.

\end{thebibliography}
\end{document}